\documentclass[12pt]{article}
\usepackage{amsthm}
\usepackage{amsmath}
\usepackage{amssymb}
\usepackage{times}
\usepackage{graphicx}
\usepackage{multirow}
\usepackage{lineno}
\usepackage{placeins}
\usepackage[utf8]{inputenc} 
\usepackage[authoryear]{natbib}
\usepackage{rotating}
\usepackage{bbm}
\usepackage{latexsym}

\newcommand{\captionfonts}{\normalsize}

\makeatletter  
\long\def\@makecaption#1#2{%
	\vskip\abovecaptionskip
	\sbox\@tempboxa{{\captionfonts #1: #2}}%
	\ifdim \wd\@tempboxa >\hsize
	{\captionfonts #1: #2\par}
	\else
	\hbox to\hsize{\hfil\box\@tempboxa\hfil}%
	\fi
	\vskip\belowcaptionskip}
\makeatother   

\usepackage{hyperref}
\usepackage[svgnames,table]{xcolor}
\usepackage{subcaption} 
\usepackage{csquotes}
\usepackage{cleveref}
\usepackage{booktabs}
\usepackage{siunitx}
\usepackage{mathrsfs} 
\usepackage{multirow}

\makeatletter  
\relax 
\providecommand\hyper@newdestlabel[2]{}
\providecommand\HyperFirstAtBeginDocument{\AtBeginDocument}
\HyperFirstAtBeginDocument{\ifx\hyper@anchor\@undefined
\global\let\oldcontentsline\contentsline
\gdef\contentsline#1#2#3#4{\oldcontentsline{#1}{#2}{#3}}
\global\let\oldnewlabel\newlabel
\gdef\newlabel#1#2{\newlabelxx{#1}#2}
\gdef\newlabelxx#1#2#3#4#5#6{\oldnewlabel{#1}{{#2}{#3}}}
\AtEndDocument{\ifx\hyper@anchor\@undefined
\let\contentsline\oldcontentsline
\let\newlabel\oldnewlabel
\fi}
\fi}
\global\let\hyper@last\relax 
\gdef\HyperFirstAtBeginDocument#1{#1}
\providecommand\HyField@AuxAddToFields[1]{}
\providecommand\HyField@AuxAddToCoFields[2]{}
\@writefile{toc}{\contentsline {section}{\numberline {A}Model parameter, benchmark function, and unabridged result tables}{1}{appendix.A}\protected@file@percent }
\@writefile{lot}{\contentsline {table}{\numberline {A1}{\ignorespaces Parameters for the model neurons and synaptic filters used in the experiments. These are the neuron and synapse model parameters used in the experiments, unless explicitly indicated otherwise.\relax }}{1}{table.caption.1}\protected@file@percent }
\providecommand*\caption@xref[2]{\@setref\relax\@undefined{#1}}
\@writefile{lot}{\contentsline {table}{\numberline {A2}{\ignorespaces Nonlinearity model parameters $b_0$, $b_1$, $b_2$, $a_0$, $a_1$, $a_2$ (\cref  {eqn:h_model_prototype}). $J_\mathrm  {max}$ and $J_\mathrm  {min}$ are the predicted absolute maximum/minimum somatic current. $\alpha $, $\beta $ characterize the rectified linear unit used as neuron nonlinearity $G[J]$ in the spike noise experiment.\relax }}{1}{table.caption.2}\protected@file@percent }
\@writefile{lot}{\contentsline {table}{\numberline {A3}{\ignorespaces Equations and contour plots detailing the multivariate functions that were analyzed in the network experiment. The functions $f(x, y)$ are scaled such that they map onto the interval $[0, 1]$ for $(x, y) \in [0, 1]^2$. Contour plots and RMS values (standard deviations) are over the input domain $[0, 1]^2$ (axis labels omitted due to space constraints); purple (dark) corresponds to a value of zero, yellow (light) to a value of one.\relax }}{2}{table.caption.3}\protected@file@percent }
\@writefile{lof}{\contentsline {figure}{\numberline {A1}{\ignorespaces Regularisation factor sweep for Experiment 2. Lines depict the median approximation error over 1000 randomly generated functions with spatial low-pass filter coefficient $\sigma = 1$. The final regularisation factors we selected are $\lambda =80$ for the current-based nonlinearity, $\lambda = 2$ for the current-based nonlinearity in a two-layer network and $\lambda = 0.01$ for the conductance-based nonlinearity.\relax }}{3}{figure.caption.4}\protected@file@percent }
\@writefile{lot}{\contentsline {table}{\numberline {A4}{\ignorespaces Spiking neural network function approximations. Error values correspond to the normalised RMSE (relative to the RMS/standard deviation of the target function). $E_\mathrm  {model}$ is the static (i.e.,\nobreakspace  {}without spike noise) decoding error assuming the dendritic and somatic models are accurate. $E_\mathrm  {net}$ is the decoding error of the target population when feeding the input through the entire network.\relax }}{4}{table.caption.5}\protected@file@percent }
\@writefile{toc}{\contentsline {section}{\numberline {B}Pre-filter experiment and results}{5}{appendix.B}\protected@file@percent }
\@writefile{lot}{\contentsline {table}{\numberline {B1}{\ignorespaces Spiking neural network function approximations \emph  {with} pre-filter. Error values correspond to the normalised RMSE (relative to the RMS/standard deviation of the target function). $E_\mathrm  {model}$ is the static (i.e.,\nobreakspace  {}without spike noise) decoding error assuming the dendritic and somatic models are accurate. $E_\mathrm  {net}$ is the decoding error of the target population when feeding the input through the entire network.\relax }}{6}{table.caption.6}\protected@file@percent }
\@writefile{lof}{\contentsline {figure}{\numberline {B1}{\ignorespaces Sweep over the regularization parameter $\sigma $ where $\lambda = (a_\mathrm  {max} \sigma )^2 = (100 \sigma )^2$ and the pre-filter time-constant $\tau $. Results are averages on a $33 \times 32$ grid with ten samples per point. Contour plots show $\qopname  \relax o{log}_{10}(E_\mathrm  {net})$ when computing multiplication $f(x, y) = x y$ over $(x, y) \in [0, 1]^2$. \textcircled {\relax \fontsize  {9}{11}\selectfont  \abovedisplayskip 8.5\p@ plus3\p@ minus4\p@ \abovedisplayshortskip \z@ plus2\p@ \belowdisplayshortskip 4\p@ plus2\p@ minus2\p@ \def \leftmargin \leftmargini \parsep 4\p@ plus2\p@ minus\p@ \topsep 8\p@ plus2\p@ minus4\p@ \itemsep 4\p@ plus2\p@ minus\p@ {\leftmargin \leftmargini \topsep 4\p@ plus2\p@ minus2\p@ \parsep 2\p@ plus\p@ minus\p@ \itemsep \parsep }\belowdisplayskip \abovedisplayskip A} corresponds to the regularization factor $\sigma $ minimizing the error for no pre-filter (as discussed in the main section). \textcircled {\relax \fontsize  {9}{11}\selectfont  \abovedisplayskip 8.5\p@ plus3\p@ minus4\p@ \abovedisplayshortskip \z@ plus2\p@ \belowdisplayshortskip 4\p@ plus2\p@ minus2\p@ \def \leftmargin \leftmargini \parsep 4\p@ plus2\p@ minus\p@ \topsep 8\p@ plus2\p@ minus4\p@ \itemsep 4\p@ plus2\p@ minus\p@ {\leftmargin \leftmargini \topsep 4\p@ plus2\p@ minus2\p@ \parsep 2\p@ plus\p@ minus\p@ \itemsep \parsep }\belowdisplayskip \abovedisplayskip B} corresponds to the $\sigma $, $\tau $ pair minimizing the error when using a pre-filter.\relax }}{7}{figure.caption.7}\protected@file@percent }
\@writefile{toc}{\contentsline {section}{\numberline {C}The two-compartment LIF model cannot compute XOR}{8}{appendix.C}\protected@file@percent }
\makeatother   

\crefformat{figure}{Fig~#2{\MakeUppercase{#1}}#3}
\Crefformat{figure}{Fig~#2{\MakeUppercase{#1}}#3}
\crefmultiformat{figure}{Figs~#2{\MakeUppercase{#1}}#3}{ and~#2{\MakeUppercase{#1}}#3}{, #2{\MakeUppercase{#1}}#3}{ and~#2{\MakeUppercase{#1}}#3}
\Crefmultiformat{figure}{Figs~#2{\MakeUppercase{#1}}#3}{ and~#2{\MakeUppercase{#1}}#3}{, #2{\MakeUppercase{#1}}#3}{ and~#2{\MakeUppercase{#1}}#3}
\crefrangeformat{figure}{Fig~#3{\MakeUppercase{#1}}#4 to~#5{\MakeUppercase{#2}}#6}
\Crefrangeformat{figure}{Fig~#3{\MakeUppercase{#1}}#4 to~#5{\MakeUppercase{#2}}#6}

\begingroup
\makeatletter
\@for\theoremstyle:=theorem,corollary,definition\do{%
	\expandafter\g@addto@macro\csname th@\theoremstyle\endcsname{%
		\addtolength\thm@preskip{1em}
		\addtolength\thm@postskip{1em}
	}%
}
\endgroup

\theoremstyle{definition}

\begin{document}

\hspace{13.9cm}1

\ \vspace{20mm}\\

{\LARGE Passive nonlinear dendritic interactions as a computational resource in spiking neural networks}

\ \\
{\bf \large Andreas Stöckel\textsuperscript{1}, Chris Eliasmith\textsuperscript{1}}\\
{\textsuperscript{1} Centre for Theoretical Neuroscience, University of Waterloo, Waterloo, Ontario, Canada}\\
%


\thispagestyle{empty}

\ \vspace{-0mm}\\
\begin{center} {\bf Abstract} \end{center}
Nonlinear interactions in the dendritic tree play a key role in neural computation. Nevertheless, modeling frameworks aimed at the construction of large-scale, functional spiking neural networks, such as the Neural Engineering Framework, tend to assume a linear superposition of post-synaptic currents.
In this paper, we present a series of extensions to the Neural Engineering Framework that facilitate the construction of networks incorporating Dale's principle and nonlinear conductance-based synapses.
We apply these extensions to a two-compartment LIF neuron that can be seen as a simple model of passive dendritic computation.
We show that it is possible to incorporate neuron models with input-dependent nonlinearities into the Neural Engineering Framework without compromising high-level function and that nonlinear post-synaptic currents can be systematically exploited to compute a wide variety of multivariate, bandlimited functions, including the Euclidean norm, controlled shunting, and non-negative multiplication.
By avoiding an additional source of spike noise, the function-approximation accuracy of a single layer of two-compartment LIF neurons is on a par with or even surpasses that of two-layer spiking neural networks up to a certain target function bandwidth.

\section{Introduction}
A central challenge in theoretical neuroscience is to describe how biological mechanisms ultimately give rise to behavior.
One way to approach this challenge is to build models of neurobiological systems that generate the behavior of interest to a researcher. Since constructing models that span multiple levels of abstraction is typically difficult, theoretical neuroscientists are working on methods that facilitate mapping high-level behaviour onto neural mechanisms.
Such modeling frameworks include the Neural Engineering Framework (NEF)~\citep{eliasmith2003neural,eliasmith2013build}, Efficient, Balanced Spiking Networks (EBN)~\citep{boerlin2011spikebased,boerlin2013predictive}, and FORCE~\citep{sussillo2009generating,nicola2017supervised}.
Generally speaking, these approaches describe how to translate dynamical systems---corresponding to some hypothesized behavioral model---into an idealized spiking neural network that adheres to the desired neurophysiological constraints, for example neural tuning, firing rate distributions, and population-level connectivity ~\citep{komer2016unified,nicola2017supervised}.
This mechanistic grounding facilitates model validation by enabling a direct comparison of simulation results and empirical data~\citep[e.g.,][]{stewart2012learning,bekolay2014,duggins2017effects,voelker2018improvinga,gosmann2020}.

The frameworks mentioned above primarily rely on two biophysical phenomena as computational primitives: synaptic filtering and the nonlinear relationship between somatic input currents and the neural response. Somatic response models range from leaky integrate-and-fire (LIF) to Hodgkin-Huxley type dynamics~\citep{schwemmer2015constructing,eliasmith2016biospaun,duggins2017incorporating}. Crucially however, these approaches typically assume that post-synaptic currents are a linear superposition of filtered pre-synaptic events. Nonlinear interactions between input channels as they may occur when modeling conductance-based synapses or dendritic structures are typically ignored.

While some research exists that explores the effect of nonlinear post-synaptic currents within these frameworks \citep{bobier2014unifying,thalmeier2016learning,alemi2018learning}, these nonlinearities are seldom systematically exploited. Yet, empirical and theoretical work suggests that active and passive nonlinear effects within the dendritic tree---and not only the soma---are at least partially responsible for the complex responses observed in some biological neurons, including cortical pyramidal cells~\citep{mel1994information,koch1999biophysics,polsky2004computational}. \cite{london2005dendritic} argue that in addition to voltage-gated ionic currents, fundamental passive effects such as shunting inhibition are worth being investigated as computational resources.

Put differently, current functional modeling frameworks only consider a subset of the computational resources available in individual neurons and thus underestimate their computational power. Modelers wishing to multiply two signals might for example be forced to introduce an additional layer of neurons, although---in biology---the interplay between excitation and inhibition within the dendrites of a single neuron layer could have the same effect \citep{koch1999biophysics}. The goal of this paper is to present mathematically tractable methods that allow researchers to take nonlinear post-synaptic currents into account. We demonstrate, as demanded by \cite{london2005dendritic}, that the interactions between passive conductance-based input channels within a single dendritic compartment provide significant computational advantages over standard LIF neurons even within a noisy spiking neural network with low firing rates and small neuron counts.

Specifically, we extend the NEF towards systematically exploiting nonlinear post-synaptic currents. The NEF has been applied to various research areas, including low-level modeling of neurobiological systems~\citep{kuo2005integrating,tripp2009search,bobier2014unifying}, and studying large-scale models of cognitive systems grounded in biology~\citep{eliasmith2012largescale,eliasmith2013build,eliasmith2016biospaun}. A software implementation of the NEF is part of the neural simulation package Nengo~\citep{bekolay2014nengo} and has been used as a neural compiler targeting analog and digital neuromorphic hardware~\citep{choudhary2012silicon,mundy2015efficient,berzish2016realtime,blouw2018benchmarking,neckar2019braindrop}.

The main contributions of this paper are as follows. First, we present a series of extensions to the NEF that improve its compatibility with more biologically detailed neuron models. We describe how to enforce nonnegative weights in conjunction with Dale's principle and extend the NEF towards nonlinear post-synaptic currents, thereby lifting some long-standing limitations of the NEF.
Second, we derive a post-synaptic current model for a two-compartment leaky integrate-and-fire (LIF) neuron that can be interpreted as a simple dendritic nonlinearity.
Third, we demonstrate that a single layer of two-compartment LIF neurons can compute a wide variety of functions with an error smaller than or on a par with the accuracy achieved by a comparable two-layer spiking neural network, as long as the target function does not surpass a certain bandwidth.

\section{Materials and Methods}
We begin with a review of relevant portions of the Neural Engineering Framework (NEF) followed by four novel extensions: decoding in current space, nonnegative weights, equality relaxation for subthreshold currents, and the input-dependent nonlinearity model $H$. These extensions facilitate incorporating biological detail into NEF models, including nonlinear, dendritic post-synaptic currents. We close with a description of the two-compartment LIF neuron, and the derivation of a corresponding nonlinearity $H$.\footnote{The methods presented in this section have been implemented as a software library extending the Nengo simulation package, see \url{https://github.com/astoeckel/nengo-bio}. A standalone version used to conduct the experiments in this paper can be found in the supplemental material.}

\subsection{The Neural Engineering Framework (NEF)}
\label{sec:nef}

At its core, the NEF describes three principles that govern the construction of models of neurobiological systems \citep{eliasmith2003neural}. These principles apply established concepts from artificial neural networks to time-continuous spiking neural networks in a way that facilitates the integration of neurophysiological constraints. The first principle postulates that populations of neurons represent values in a relatively low-dimensional manifold within their high-dimensional activity space via nonlinear encoding and linear decoding.  According to the second principle, network connectivity defines transformations as mathematical functions of the represented values. Finally, the third principle states that recurrent connections approximate arbitrary dynamical systems in which the represented values are state variables. Although we will not discuss recurrent networks in this paper, it should be noted that the methods presented here are fully compatible with the third principle and as such can be used to implement dynamical systems.

\subsubsection*{Representation}
\label{sec:nef_representation}

A fundamental assumption of the NEF is that populations of spiking neurons represent $d$-di\-men\-sion\-al vectors $\vec x \in \mathbb{R}^d$ by means of nonlinear encoding and linear decoding.
In essence, each neuron population is treated as a single hidden layer neural network with linear input and output units. Similar schemes have been proposed since the late 1960s, an early example being Marr's model of pattern extraction in the cerebellar granule layer \citep{marr1969theory}.

\begin{figure}[p]
	\centering%
	\includegraphics[scale=0.9]{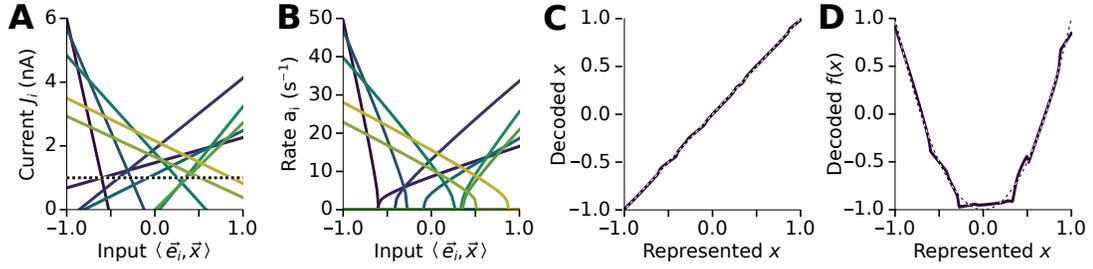}
	\hspace{0.25cm}
	{\phantomsubcaption\label{fig:nef_overview_a}}
	{\phantomsubcaption\label{fig:nef_overview_b}}
	{\phantomsubcaption\label{fig:nef_overview_c}}
	{\phantomsubcaption\label{fig:nef_overview_d}}
	\caption{Representation and transformation in the Neural Engineering Framework. Example for a population of $n = 10$ neurons. (A) Randomly selected linear~(affine) current translation functions $J_i(\xi) = \alpha_i \xi + \beta_i$ map the encoded value $\xi = \langle \vec x, \vec e_i \rangle$ onto an input current $J_i$. The dotted line corresponds to the spike threshold $J_\mathrm{th}$. (B) Tuning curves for each neuron in the population after applying the somatic nonlinearity $G[J]$~(eq.~\ref{eqn:lif_rate}). (C) Reconstructing the represented value from neuron activity by means of linear decoding. (D) Approximating the function $f(x) = 2x^2 - 1$ with a different linear decoding. Dashed lines correspond to the mathematical function being approximated.}
	\label{fig:nef_overview}
\end{figure}

For the encoding process, each neuron $i$ in a population of size $n$ is assigned a tuning-curve $a_i(\vec x)$ that maps any represented value $\vec x$ onto a corresponding activity~(\cref{fig:nef_overview_a,fig:nef_overview_b}). Mathematically,
\begin{linenomath*}\begin{align}
a_i(\vec x) = G\big[J_i\big(\langle \vec x, \vec e_i \rangle\big)\big] = G\big[\alpha_i\langle \vec x, \vec e_i \rangle + \beta_i\big] \,,
\label{eqn:encoding}
\end{align}\end{linenomath*}
where the encoding vector $\vec e_i$ projects the input $\vec x$ onto a scalar $\xi$, which in turn is translated by $J_i(\xi)$ into a somatic current. While the specific current translation function $J_i(\xi)$ depends on modeling assumptions, it is often defined as a first-order polynomial parametrized by a gain $\alpha_i$ and a bias current $\beta_i$. When building models, encoders $\vec e_i$, biases $\beta_i$, and gains $\alpha_i$, must be selected in such a way that the population exhibits diverse neural tuning, while at the same time staying within modeling constraints such as maximum firing rates for the range of represented $\vec x$. The nonlinear neuron response $G[J]$ defines the spiking neuron model used in the network. For a wide variety of neuron models, this can be characterized by a rate approximation that maps the somatic input current $J$ onto an average firing rate.

Critically, this mapping does not suggest the adoption of a \enquote{rate code} by the NEF, but rather is a convenience for solving the synaptic weight optimization problem. That is, \cref{eqn:encoding} is merely normative: it defines the average activity that a modeler \emph{expects} individual neurons to have when the population represents a certain $\vec x$. This normative constraint can be dropped entirely, and all optimization done in the spiking domain within the NEF, but the computational costs are significantly higher~\citep{macneil2011finetuning,eliasmith2013build}. In this paper, we purely focus on offline optimization of the synaptic weights. Hence, the methods we present should in no way be interpreted as a theory of learning within nervous systems, but purely as a method for constructing models of mature, already trained, neural systems.

Notice that for the purpose of modelling neurobiological systems, all functions and parameters listed above can be hand-tuned by the modeler to match neurophysiological data. For example, as explored in \cite{eliasmith2003neural}, two-dimensional encoding vectors $\vec e_i$ can be used to reproduce a wide variety of localised tuning curves found in the neuroscience literature. Since NEF models, in contrast to classical multi-layer neural networks, are typically not globally optimized, tuning curves are maintained throughout the weight optimization process and neural activities can be interpreted at any point in the simulation. Global stochastic gradient descent can be applied to parts of the network \citep{hunsberger2015spiking} at the cost of negatively impacting the interpretability and biologically informed tuning of individual neuron populations.

Complementary to the encoding operation is decoding, which reconstructs an approximation of the represented value from the momentary population activity at time $t$ by multiplication with a decoding matrix $D \in \mathbb{R}^{d \times n}$, i.e., $\vec x(t) \approx D \vec a(t)$~(\cref{fig:nef_overview_c}). Finding $D$ can be phrased as a Tikhonov regularized least-squares optimization problem
\begin{linenomath*}\begin{align}
\min_D \sum_{k = 1}^N \| D \vec a(\vec x_k) - \vec x_k \|^2_2  + \lambda N \| D \|^2_\mathrm{F}
&= \min_{D} \| DA - X \|_\mathrm{F}^2 + \lambda N \| D \|_\mathrm{F}^2 \,,
\label{eqn:decoding}
\end{align}\end{linenomath*}
where $\|\cdot\|_2$ is the $L_2$-norm, $\|\cdot\|_\mathrm{F}$ is the Frobenius matrix norm, $\vec x_k$ is one of $N$ training samples, $A \in \mathbb{R}^{n \times N}$ is a matrix of population responses for each sample, $X \in \mathbb{R}^{d \times N}$ is a matrix of training samples, and $\lambda$ is a regularization term accounting for spike variability and other sources of error \citep{eliasmith2003neural}.

The decoders $D$ are usually optimized offline under the assumption of the rate model $G[J]$. As formulated, the same $D$ can be used to decode represented values through time in spiking networks when defining activity $\vec a(t)$ as low-pass filtered population spike trains. Linear low-pass filters are a common model for post-synaptic currents \citep{roth2009modeling} and usually employed in spiking NEF networks \citep{eliasmith2003neural}.

\subsubsection*{Transformation}
\label{sec:nef_transformation}

Nonlinear encoding and linear decoding schemes similar to the one described above have been analyzed in more detail by researchers in the field of machine learning \citep{broomhead1988radial}. Neuron population tuning-curves $\vec a(\vec x)$ span a function space with a set of non-orthogonal basis functions. We can approximate any continuous function over the represented values to a desired degree by linearly combining a finite number of these basis functions~\citep{hornik1989multilayer}.
Specifically, the linear projection $D$ in \cref{eqn:decoding} approximates the identity function. By modifying the loss function and substituting $X$ with a matrix $X^f$ of target vectors $f(\vec x_k)$ we can solve for decoders $D^f$ approximating some function $f$~(\cref{fig:nef_overview_d})
\begin{linenomath*}\begin{align}
\min_{D^f} \sum_{k = 1}^N \frac{1}2 \| D^f \vec a(\vec x_k) - f(\vec x_k) \|^2_2  + \lambda N \| D^f \|^2_\mathrm{F} \,.
\label{eqn:decoding_f}
\end{align}\end{linenomath*}

In order to construct neural \emph{networks}, we need to find synaptic weight matrices $W \in \mathbb{R}^{m \times n}$ that connect from a pre-population of $n$ neurons to a post-population of $m$ neurons. In particular, we would like to find a $W$ that implicitly decodes $f(\vec x) = \vec y$ from the pre-population and at the same time encodes $\vec y$ in the post-population. If we assume that the current translation function $J_i(x)$ is an intrinsic part of the neuron model, or, put differently, each neuron $i$ is assigned its own response-curve $G_i[\langle \vec e_i,  \vec x\rangle] = G[J_i(\langle \vec e_i, \vec x)\rangle]$, both the encoding and decoding process are linear operators. With this simplification, $W = E D^f$ fulfils the properties listed above, where $E \in \mathbb{R}^{m \times d}$ is a matrix of post-population encoding vectors $\vec e_i$, and $D^f$ the desired function decoder for the pre-population~\citep{eliasmith2003neural}.

\begin{figure}[p]
	\centering
	\includegraphics[scale=0.9]{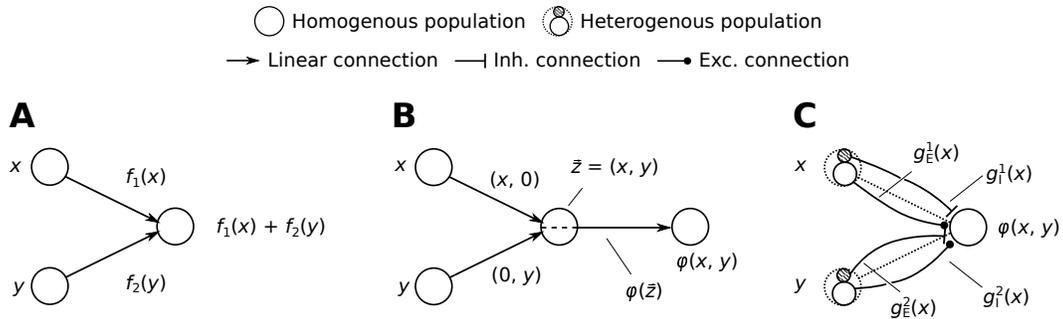}
	\vspace{0.25cm}
	{\phantomsubcaption\label{fig:network_a}}
	{\phantomsubcaption\label{fig:network_b}}
	{\phantomsubcaption\label{fig:network_c}}
	\caption{
		Multivariate computation in the NEF.
		(A) In case the current translation function $J_i$ is a part of the individual neuron response curve, NEF networks are additive: summation in activity space corresponds to addition in representational space.
		(B) Computing nonlinear multivariate functions $\varphi$ generally requires all variables to be represented in an intermediate population.
		(C) The dendritic computation scheme discussed in this paper. Two pre-populations project onto a post-population with separate excitatory and inhibitory input channels. The nonlinear interaction between these channels is exploited to compute $\varphi$.
	}
	\label{fig:network}
\end{figure}

Crucially, when combining the input from multiple pre-populations, the post-po\-pu\-la\-tion will always represent a linear combination of potentially nonlinear functions. To see this, consider two pre-populations projecting onto a common post-population, where the first projection approximates a function $f_1$, and the second a function $f_2$. Let $i$ be the index of a neuron in the post-population. Then, it holds
\begin{linenomath*}\begin{align}
\begin{aligned}
G_i\big[\big\langle \vec w^{f_1}_i, \vec a^\mathrm{pre}(\vec x_1) \big\rangle + \big\langle \vec w^{f_2}_i, \vec a^\mathrm{pre}(\vec x_2)\big\rangle\big]
&= G_i\big[\big\langle \vec e_i, D^{f_1} \vec a^\mathrm{pre}(\vec x_1) + D^{f_2} \vec a^\mathrm{pre}(\vec x_2) \big\rangle\big]\\
&\approx a^\mathrm{post}_i\big(f_1(\vec x_1) + f_2(\vec x_2)\big) \,.
\end{aligned}
\end{align}\end{linenomath*}
As a consequence, if we, for example, try to multiply two scalars $x$ and $y$, these values must be represented as a two-dimensional vector in a common pre-population. We alleviate this restriction by introducing input-dependent nonlinearities (\cref{fig:network_b,fig:network_c}).

\subsection{Extending the Neural Engineering Framework}
\label{sec:nef_ext}

The NEF as described above makes several assumptions that are not biophysically plausible. This includes the presence of a bias current $\beta_i$ in each neuron, and excitatory and inhibitory connections not being treated separately. We describe two extensions that lift the aforementioned assumptions and present an alternative synaptic weight solving procedure that takes subthreshold currents into account, followed by our proposed method for accounting for input-dependent nonlinearities.

\subsubsection*{Decoding the current translation function}
\label{sec:nef_decode_current}

In the previous subsection we assumed that the current translation function $J_i(x)$ is an intrinsic part of the neuron model. Consequently, each neuron is not only assigned a neuron-specific tuning-curve $a_i(\vec x)$, but also a neuron-specific response-curve $G_i[J]$. This comes at the cost of biological plausibility, since neurons in general do not possess a strong bias current source.

\cite{tripp2007neural} demonstrate that it is possible to robustly solve for synaptic weight matrices that approximate arbitrary post-synaptic current functions. We can use this insight to directly solve for weights that approximate the current translation function $J_i(\langle \vec e_i, \vec x \rangle)$, instead of optimizing with respect to represented values $\vec x$. Since, for now, we assume that the post-synaptic current is linear in the pre-population activities, we must find a weight vector $\vec w_i$ such that the following loss is minimized
\begin{linenomath*}\begin{align}
\min_{\vec w_i} \sum_{k=1}^N \left( J_i\big(\langle \vec e_i, f(\vec x_k)\rangle\big) - \big\langle \vec w_i, \vec a^\mathrm{pre}(\vec x_k) \rangle \right)^2 + \lambda N \| \vec w_i \|_\mathrm{2}^2 \,.
\label{eqn:decode_current}
\end{align}\end{linenomath*}
This equation can be brought into canonical least squares form and solved as before.

\subsubsection*{Nonnegative weights and Dale's principle}
\label{sec:nef_nonneg}

So far we have assumed that synaptic weights are real-valued. This is problematic for two reasons. First, the least-squares optimization proposed above arbitrarily assigns algebraic signs to the synaptic weights; we cannot specify which pre-neurons are excitatory, and which inhibitory. Being able to do so is important, since neurons tend to follow Dale's principle---individual pre-neurons exclusively influence all their post-neurons in either an excitatory or inhibitory manner~\citep{strata1999dale}. Empirical data suggest that, depending on the modeled brain region, excitatory cells outnumber inhibitory cells by a factor between two and four~\citep{hendry1981sizes,gabbott1986quantitative}.
Second, real-valued weights do not generalize to conductance-based synapses. The concept of negative conductances---in contrast to negative currents---is neither physically nor biologically plausible. Biological correlates of synaptic weights, such as the number of vesicles released from the pre-synapse or the channel density in the post-synapse, are inherently nonnegative quantities~\citep{roth2009modeling}.

An extension to the NEF that takes these biological constraints into account has been proposed by \cite{parisien2008solving}. The \emph{Parisien transform} splits each projection into an excitatory and inhibitory path, where the latter mediates the signal from the excitatory pre-population through a population of inhibitory interneurons. The solution we discuss here does not introduce interneurons, and as such does not affect the structure of the network. Modelers using the following method will have to explicitly specify inhibitory interneuron populations if present in the modeled circuit.\footnote{Our software library \texttt{nengo-bio} facilitates the process of specifying network topologies with excitatory and inhibitory populations. The library provides special syntactic sugar for networks with inhibitory interneurons built without the aforementioned Parisien transform.}

For the sake of simplicity, we assume in this paper that each population is arbitrarily split into a group of excitatory and inhibitory neurons. We write the somatic input current of post-neuron $i$ in response to pre-synaptic activity as $\langle \vec w_i^+, \vec a^+ \rangle - \langle \vec w_i^-, \vec a^- \rangle$, where, $\vec w_i^\pm$ are the nonnegative excitatory and inhibitory weight vectors and $\vec a^\pm$ the activities of the excitatory and inhibitory neurons in the pre-population. Combining this current term with \cref{eqn:decode_current} yields an optimization problem that allows us to solve for weights approximating $f$ for each individual post-neuron $i$
\begin{linenomath*}\begin{align}
\min_{\vec w_i^+, \vec w_i^-} &~\hphantom{=\,} \frac{1}2 \sum_{k=1}^N \big\| \langle \vec w_i^+,\vec a_k^+\rangle - \langle \vec w_i^-,\vec a_k^-\rangle - J_i\big(\langle \vec e_i, f(\vec x_k)\rangle\big) \big\|_2^2 + \lambda N \big\|\vec w_i^+\big\|_2^2 + \lambda N \big\|\vec w_i^- \big\|_2^2 \notag\\
&= \frac{1}2 \big\| \vec w_i'\!A' - \vec \jmath \big\|_2^2 + \lambda N \big\|\vec w_i'\big\|_2^2 \,, \,\,\text{where } \vec w_i' = (\vec w_i^+, \vec w_i^-\big),\, A' =  (A^+, -A^-\big)^T,\notag\\&~ \hspace{5.4cm} \text{and } (\vec \jmath\,)_k = J_i\big(\langle \vec e_i, f(\vec x_k)\rangle\big) \,,
\label{eqn:decode_nonneg}
\end{align}\end{linenomath*}
subject to $\vec w_i^+ \geq 0, \vec w_i^- \geq 0$. This problem is in canonical least-squares form and, as before, can be solved with a standard regularized nonnegative least-squares solver.

Alternatively, \cref{eqn:decode_nonneg} can be phrased as a convex quadratic program (QP), a generalization of least-squares optimization \citep{wright1999numerical}
\begin{linenomath*}\begin{align}
\min_{\vec w_i} &~ \frac{1}2 (\vec w_i')^T \big((A')^T A'\big) \vec w_i' - \vec\jmath^{\,T} A'  \vec w_i' +  \lambda N \| \vec w_i' \|_2^2 \,, &\text{subject to} &~ \vec w_i' \geq 0 \,.
\label{eqn:decode_nonneg_qp}
\end{align}\end{linenomath*}
The global minimum of a convex QP can be computed in polynomial time \citep{kozlov1980polynomial}. We propose a QP similar to \cref{eqn:decode_nonneg_qp} to solve for conductance-based synaptic weights in the context of the two-compartment neuron model discussed below.

\subsubsection*{Equality relaxation for subthreshold currents}
\label{sec:nef_zero_rate}

Most neurons possess a threshold current $J_\mathrm{th}$ below which their firing rate is zero. However, we do not take this into account when solving for somatic currents in \cref{eqn:decode_current,eqn:decode_nonneg,eqn:decode_nonneg_qp}---we optimize for synaptic weights that \emph{precisely} evoke certain post-synaptic currents, despite the magnitude of currents smaller than $J_\mathrm{th}$ having no effect on the neural activity in a steady-state neural response model. Instead of optimizing for equality, i.e.,~enforcing that the decoded current $J_\mathrm{dec}$ must equal the target current $J_\mathrm{tar}$, we could relax this condition to an inequality constraint $J_\mathrm{dec} \leq J_\mathrm{tar}$ whenever $J_\mathrm{tar} < J_\mathrm{th}$.

We define a new optimization problem based on \cref{eqn:decode_current} that treats target currents smaller than $J_\mathrm{th}$ as an inequality constraint
\begin{linenomath*}\begin{align}
\begin{aligned}
\min_{\vec w_i} &~\sum_{k=1}^N \frac{1}2 \mathcal{E}\left(J_i\big(\langle \vec e_i, f(\vec x_k)\rangle\big), \big\langle \vec w_i, \vec a(\vec x_k) \big\rangle\right)^2 + \lambda N \| \vec w_i \|_2^2 \,,\\
\text{where} &~ \mathcal{E}(J_\mathrm{tar}, J_\mathrm{dec}) = \begin{cases}
0 & \text{if } J_\mathrm{tar} < J_\mathrm{th} \text{ and } J_\mathrm{dec} < J_\mathrm{th} \,,\\
J_\mathrm{th} - J_\mathrm{dec} & \text{if } J_\mathrm{tar} < J_\mathrm{th} \text{ and } J_\mathrm{dec} > J_\mathrm{th} \,,\\
J_\mathrm{tar} - J_\mathrm{dec} & \text{otherwise} \,,\\
\end{cases}
\end{aligned}
\label{eqn:decode_current_subthreshold}
\end{align}\end{linenomath*}
and $N$ is the number of samples, $J_i$ the current translation function of the $i$th post-neuron, $\vec e_i$ the encoding vector, $f$ the desired target function, $\vec a(\vec x_k)$ is the pre-population activity for the $k$th sample point $\vec x_k$, $\lambda$ is the regularization factor, and $J_\mathrm{th}$ the aforementioned neuronal threshold current.

By splitting the matrix of pre-population activities $A$, and the vector of target currents $\vec \jmath$, we can rewrite \cref{eqn:decode_current_subthreshold} as a quadratic program. Let $A^+$ and $\vec \jmath^+$ be the pre-population activities and target currents for superthreshold samples, and $A^-$ be the activities for subthreshold samples. Then, the QP is given as
\begin{linenomath*}\begin{align}
\begin{aligned}
\min_{\vec w_i, \vec s_i} &~
\frac{1}{2} \vec w_i^T \big((A^+)^T A^+\big) \vec w_i  -\big( (A^+)^T (\vec \jmath^+) \big)  \vec w_i + \| \vec s_i \|_2^2 + \lambda N \| \vec w_i \|_2^2 \,,
\end{aligned}
\end{align}\end{linenomath*}
subject to $A^- \vec w_i - \vec s_i \leq J_\mathrm{th}$, where $\vec s_i$ is a vector of slack variables. Of course, the nonnegativity constraint from \cref{eqn:decode_nonneg_qp} can be incorporated into to this quadratic program.

\subsubsection*{Extension towards input-dependent nonlinearities}
\label{sec:nef_nonlinear}

\begin{figure}[p]
	\captionsetup[subfigure]{oneside,margin={0.75cm,0cm}}%
	\centering%
	\vspace{5cm}
	\includegraphics[scale=0.9]{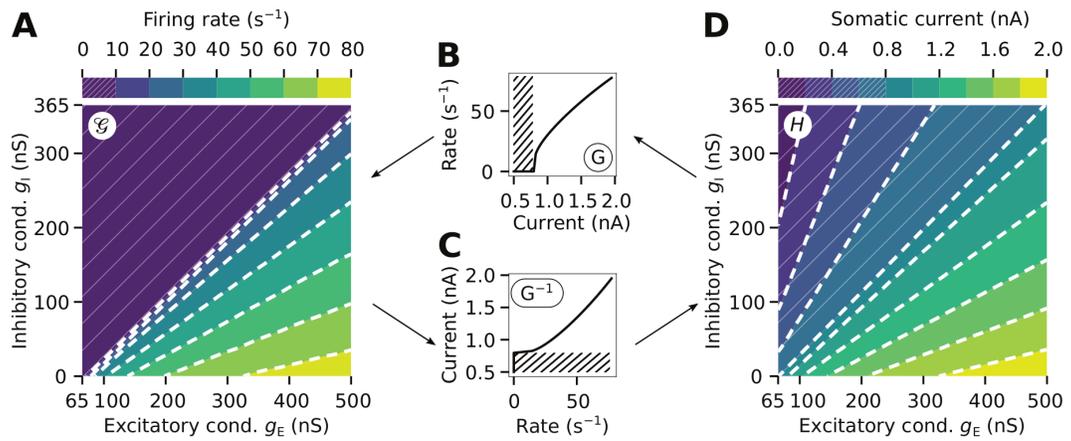}
	\vspace{0.25cm}
	\caption{Neural response curve decomposition. (A) Illustration of the multivariate neuron response curve $\mathscr{G}(g_\mathrm{E}, g_\mathrm{I})$ for a two-compartment LIF neuron with excitatory and inhibitory conductance-based channels. (B,C) The chosen somatic nonlinearity $G$ and its inverse $G^{-1}$. (D) corresponding input-dependent nonlinearity $H$. The neuron does not fire in the hatched regions, that is, $G^{-1}$ is ill-defined.} 
	\vspace{5cm}
	\label{fig:two_compartment_response_curve}
\end{figure}

Up to this point we assumed current-based synapses. Their defining property is that the somatic current $J$ is linear in the synaptic weights $\vec w$ and the pre-synaptic activities $\vec a$.
Now, consider a neuron model with $\ell$ nonlinear input channels. We write the corresponding response curve $\mathscr{G}$ (mapping from neural input onto an average firing rate) for the $i$th neuron in a population as a multivariate function
\begin{linenomath*}\begin{align}
	a_i = \mathscr{G}\big[g^1_i, \ldots, g^\ell_i\big] = \mathscr{G}\big[\langle \vec a^1_i, \vec w^1_i \rangle, \ldots, \langle \vec a^\ell_i, \vec w^\ell_i \rangle\big]\,,
	\label{eqn:rate_nonlinear}
\end{align}\end{linenomath*}
where $g^1_i, \ldots, g^\ell_i$ describe some abstract \enquote{input state}, such as the conductance of each synaptic channel in a neuron model with conductance-based synapses. As expressed in the above equation, we assume that on average each $g^k_i$ is linear in the pre-synaptic activities $\vec a^k_i$. However, we do not make any assumption regarding the effect of $g^k_i$ on the somatic current; more fundamentally, we do not assume that there exists an easily identifiable somatic current in the model at all.

The crucial idea is to mathematically reintroduce a \enquote{virtual} somatic current $J$ by decomposing $\mathscr{G}$ into an input-dependent nonlinearity $H$ and a somatic nonlineartiy $G$. We define $G$ and $H$ according to the following equivalence relations
\begin{linenomath*}\begin{align}
	                \mathscr{G}\big[g^1_i, \ldots, g^\ell_i\big] \hspace{-0.025cm}= \hspace{-0.025cm}G\big[H(g^1_i, \ldots, g^\ell_i)\big]
	\Leftrightarrow H\big(g^1_i, \ldots, g^\ell_i\big) \hspace{-0.025cm}=\hspace{-0.025cm} J
	\Leftrightarrow G\big[J\big] \hspace{-0.025cm}=\hspace{-0.025cm} \mathscr{G}\big[g^1_i, \ldots, g^\ell_i\big] \,,
	\label{eqn:def_h}
\end{align}\end{linenomath*}
where $H$ maps from the input channel state onto an average somatic current and, as before, $G$ maps from a somatic current onto the output activity. In other words, $H$ summarizes nonlinear effects caused by dendrites or conductance-based synapses.

While this formalization does not constrain $G$ and $H$ beyond the above equivalence, a sensible choice for $G$ and $H$ will facilitate solving for synaptic weights. For example, if the neuron model in question is an extension to the current-based LIF neuron model, it makes sense to select $G$ as the standard LIF response curve. Then, $H$ translates the input state into an \enquote{LIF-equivalent somatic current}.

As we show in the next section, $H$, or at least a parametrized surrogate for $H$, can be derived in closed form in some cases. In case this is not feasible, it is possible to purely rely on empirical data for $H$ by sampling the neuron output rate over varying synaptic states. Assume that we can only observe $H$ indirectly by controlling the input channels of our model neuron and measuring the output activity $\mathscr{G}$. Depending on our choice of $G[J]$, we can apply an inverse mapping $G^{-1}$ to the recorded data to obtain $H$. See \cref{fig:two_compartment_response_curve} for an illustration.

When solving for weights that approximate a specific function, recall from the above review that the first NEF principle normatively assigns a somatic current $J_i(\vec x)$ to each post-neuron $i$ and representational state $\vec x$. Correspondingly, given $H$, we can combine \cref{eqn:rate_nonlinear} with the current-decoding problem discussed above, as well as the nonnegativity constraint and the equality relaxation:
\begin{linenomath*}\begin{align}
\min_{\vec w^1_i, \ldots, \vec w^\ell_i} &~
\sum_{k = 1}^N \mathcal{E}\left(
J_i\big(\langle \vec e_i, f(\vec x_k) \rangle\big),
H\big(\langle \vec w^1_i, \vec a^1_k \rangle, \ldots, \langle \vec w^\ell_i, \vec a^\ell_k \rangle\big)
\right)^2 + \lambda N \sum_{j = 1}^\ell \| \vec w^j_i \|_2^2\,,
\label{eqn:decode_nonlinear_synapses}
\end{align}\end{linenomath*}
subject to $\vec w_i^j \geq 0$, where $\mathcal{E}$ is as defined in \cref{eqn:decode_current_subthreshold}.

\subsection{Two-compartment leaky integrate-and-fire neuron model}
\label{sec:model}

In the previous section we established the abstract notion of an input-depdendent nonlinearity $H$. We now derive a concrete $H$ for a biologically plausible extension to the leaky integrate-and-fire (LIF) neuron with nonlinear conductance-based synapses. In particular, we consider a two-compartment version of the LIF neuron originally described in \cite{vu1993mechanism} and subsequently discussed by \cite{koch1999biophysics} and \cite{capaday2006direct}. In addition to the active, somatic compartment, the two-compartment LIF model possesses a resistively coupled passive compartment that represents excitatory and inhibitory input into the dendritic tree. Depending on the coupling conductance $g_\mathrm{C}$, the input may either be interpreted as distal or proximal.

We first review the neuron model itself, derive a parametrized surrogate for the nonlinearity $H$, and finally propose a convex quadratic program for this $H$ that can be used to solve for synaptic weights.

\subsubsection*{Model description}
\label{sec:model_description}

\begin{figure}[t]
	\centering
	\vspace{0.75cm}
	\includegraphics[scale=0.9]{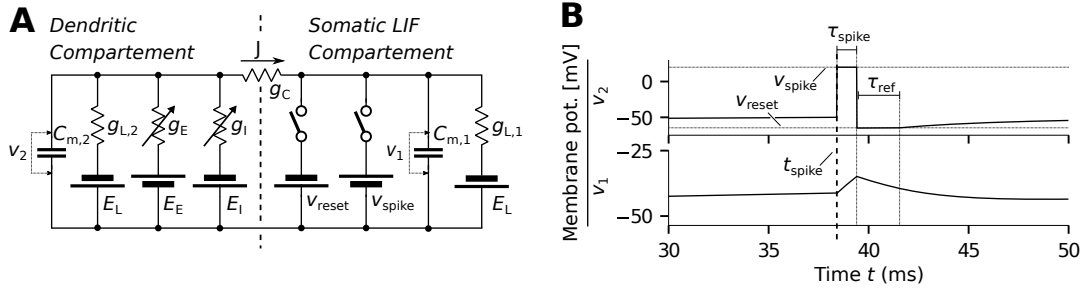}
	{\phantomsubcaption\label{fig:neuron_model}}
	{\phantomsubcaption\label{fig:trace}}
	\vspace{0.75cm}
	\caption{Two-compartment LIF neuron model. (A) Equivalent circuit diagram of the two-compartment LIF neuron model. The \enquote{somatic} compartment corresponds to a classical LIF neuron; the \enquote{dendritic} compartment is resistively coupled to the somatic compartment. The current flowing between the two compartments corresponds to the somatic current $J = H(g_\mathrm{E}, g_\mathrm{I})$. (B) Voltage trace for the two-compartment model state variables $v_1$ and $v_2$ during spike production. The artificial depolarization following spike events in the somatic compartment significantly affects $v_2$.}
\end{figure}

The sub-threshold dynamics of the conductance-based two-compartment LIF model can be expressed as a two-dimensional system of linear differential equations
\begin{linenomath*}\begin{align}
\begin{aligned}
\frac{d}{dt} C_{\mathrm{m},1} v_1 &=
g_\mathrm{C} (v_2 - v_1)
+ g_\mathrm{L,1} (E_\mathrm{L} - v_1) \,, \\
\frac{d}{dt} C_{\mathrm{m},2} v_2 &=
g_\mathrm{C} (v_1 - v_2)
+ g_\mathrm{L,2}  (E_\mathrm{L} - v_2)
+ g_\mathrm{E}(t) (E_\mathrm{E} - v_2)
+ g_\mathrm{I}(t) (E_\mathrm{I} - v_2) \,,
\end{aligned}
\label{eqn:neuron_model}
\end{align}\end{linenomath*}
where state variables $v_1$, $v_2$ correspond to the membrane potential of the active somatic compartment and the passive dendritic compartment, respectively. $C_{\mathrm{m},1}$, $C_{\mathrm{m},2}$ are the compartment capacitances, $g_\mathrm{C}$ is the inter-compartment coupling conductance, $g_{\mathrm{L},1}$, $g_{\mathrm{L},2}$ are the individual compartment leak conductances, $g_\mathrm{E}(t)$, $g_\mathrm{I}(t)$ are the momentary excitatory and inhibitory conductances of the dendritic compartment as evoked by pre-synaptic spikes, and $E_\mathrm{L}$, $E_\mathrm{E}$, $E_\mathrm{I}$ are the conductance-channel reversal potentials. An equivalent circuit diagram of the model is shown in \cref{fig:neuron_model}.

In contrast to point-neuron models, and as pointed out by \cite{capaday2006direct}, a multi-compartment model mandates an explicit spike model. The strong depolarization in the somatic compartment during spike production propagates backwards into the denritic compartment and has a significant effect on its membrane potential (\cref{fig:trace}). The model accounts for this with a \enquote{spike-phase} occurring right before the LIF refractory period. The spike phase is implemented by clamping the somatic compartment to a voltage $v_\mathrm{spike}$ over a time $\tau_\mathrm{spike}$ whenever the somatic membrane potential crosses the threshold $v_\mathrm{th}$. Subsequently, the membrane is clamped to $v_\mathrm{reset}$ for a period $\tau_\mathrm{ref}$.

\subsubsection*{Somatic current surrogate model}
\label{sec:somatic_current_model}

We assume that $H(g_\mathrm{E}, g_\mathrm{I})$ is equivalent to the current flowing from the dendritic into the somatic compartment (\cref{fig:neuron_model}), that is $H(g_\mathrm{E}, g_\mathrm{I}) = g_\mathrm{C} (v_2 - v_1)$. Considering the definition in \cref{eqn:def_h}, this implies that $G$ is the standard LIF response curve
\begin{linenomath*}\begin{align}
G[J] = \left( {\tau_\mathrm{ref} + \tau_\mathrm{spike} - \frac{C_\mathrm{m}}{g_\mathrm{L}} \ln \left(1 - \frac{(v_\mathrm{th} - E_\mathrm{L}) g_\mathrm{L}}{J} \right)}  \right)^{-1}\,,
\label{eqn:lif_rate}
\end{align}\end{linenomath*}
where $\tau_\mathrm{ref}$, $\tau_\mathrm{spike}$ are the length of the refractory and spike periods~(\cref{fig:trace}), $C_\mathrm{m}$ is the membrane capacitance, $v_\mathrm{th}$ is the threshold potential, $E_\mathrm{L}$ is the leak reversal or resting potential, and $g_\mathrm{L}$ is the leak conductance \citep{eliasmith2003neural}. Yet, in practice, as pointed out by \cite{hunsberger2014competing}, and as we demonstrate in our experiments below, a rectified linear unit (ReLU) may be a sensible choice for $G$ as well when modeling noisy input.

Unfortunately, when considering both sub- and superthreshold dynamics, there exists no exact closed-form solution for the average somatic current given constant $g_\mathrm{E}$, $g_\mathrm{I}$. Instead, our approach is to select a parametrized surrogate model for $H$ and to fit the model parameters to results obtained from numerical simulations to account for inaccuracies in our derivation.

Assuming the subthreshold dynamics described in \cref{eqn:neuron_model} are in equilibrium, $g_\mathrm{E}$, $g_\mathrm{I}$ are constant, and applying the above definition of $H(g_\mathrm{E}, g_\mathrm{I})$, we get
\begin{linenomath*}\begin{align}
H(g_\mathrm{E}, g_\mathrm{I}) = g_\mathrm{C} (v_2 - v_1) =
g_\mathrm{C} \frac{  g_{\mathrm{L},2} (E_\mathrm{L} - v_1)
	+ g_\mathrm{E}   (E_\mathrm{E} - v_1)
	+ g_\mathrm{I}   (E_\mathrm{I} - v_1) }
{g_\mathrm{C} + g_{\mathrm{L},2} + g_\mathrm{E} + g_\mathrm{I}} \,.
\label{eqn:h_model_prototype}
\end{align}\end{linenomath*}
A single-compartment model can be derived by taking the limit of this equation for $g_\mathrm{C} \to \infty$. In this case, and as demonstrated in \cite{stockel2017point}, $H$ is an affine function and less interesting from a computational perspective. In general, we expect $H$ to become more nonlinear as $g_\mathrm{C}$ decreases. While a higher degree of nonlinearity may be desirable for computation in a neural network, $g_\mathrm{C}$ cannot be made arbitrarily small, as this limits the maximum somatic input current (see below).

\begin{figure}
	\centering
	\includegraphics[scale=0.9]{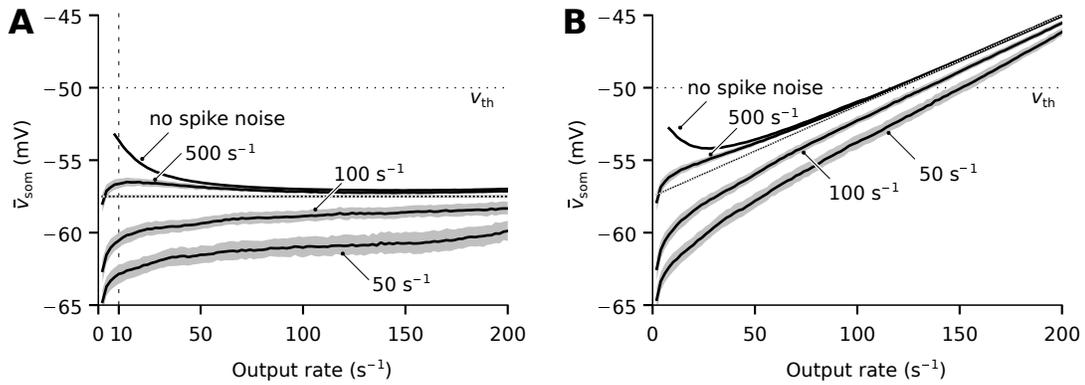}
	{\phantomsubcaption\label{fig:avg_vsom_no_ref}}
	{\phantomsubcaption\label{fig:avg_vsom_with_ref}}
	\vspace{0.25cm}
	\caption{Average LIF membrane potential over firing rate. Data corresponds to a standard LIF neuron (i.e.,~the somatic compartment) over varying output rates. Data measured by capturing \num{1000} membrane potential traces for a ten second current sweep with superimposed spike noise from a Poisson spike source. Individual lines correspond to the effective rate of the spike source, where smaller input rates are equivalent to a larger amount of noise on the input current. Shaded areas correspond to the 25/75 percentile. (A) Average potential excluding the refractory period and spike production. Except for very low firing rates, the average potential remains relatively constant. Dotted line corresponds to ${\overline v}_\mathrm{som} = \frac{1}2 (v_\mathrm{reset} + v_\mathrm{th})$. (B) Average potential including the refractory and spike production. Dotted line corresponds to a linear model that takes the relative amount of time spent in the subthreshold, spike, and refractory regime into account.}
	\label{fig:avg_vsom}
\end{figure}

Since the somatic membrane potential $v_1$ is clamped during the spike and refractory phases, the current flowing into the soma during those times does not influence the overall firing rate (ignoring the feedback effect on $v_2$ discussed earlier). Once the neuron is tonically firing, the somatic membrane potential oscillates between $v_\mathrm{reset}$ and $v_\mathrm{th}$. We can thus substitute $v_1$ with a constant average membrane potential ${\overline v}_\mathrm{som} \approx \frac{1}2 (v_\mathrm{reset} + v_\mathrm{th})$ (\cref{fig:avg_vsom}). Under these assumptions, $H(g_\mathrm{E}, g_\mathrm{I})$ from \cref{eqn:h_model_prototype} can be written as a parametrized rational function%
\begin{linenomath*}\begin{align}
H(g_\mathrm{E}, g_\mathrm{I}) &= \frac{b_0 + b_1 g_\mathrm{E} + b_2 g_\mathrm{I}}
{a_0 + a_1 g_\mathrm{E} + a_2 g_\mathrm{I}} \,,
& \text{where } a_0, a_1, a_2, g_\mathrm{E}, g_\mathrm{I} \geq 0 \,.
\label{eqn:h_model}
\end{align}\end{linenomath*}
This equation has one superfluous degree of freedom in the parameter space. Setting $b_1 = 1$ tends to be a numerically stable normalization.

\Cref{eqn:h_model} implies an absolute maximum and minimum somatic current; $H(g_\mathrm{E}, g_\mathrm{I})$ maps onto an open interval $(J_\mathrm{max}, J_\mathrm{min})$, where
\begin{linenomath*}\begin{align}
\begin{aligned}
J_\mathrm{min} &= \lim_{g_\mathrm{I} \to \infty} H(g_\mathrm{E}, g_\mathrm{I}) = -\frac{b_2}{a_2} = g_\mathrm{C} (E_\mathrm{I} - {\overline v}_\mathrm{som}) \,,\\
J_\mathrm{max} &= \lim_{g_\mathrm{E} \to \infty} H(g_\mathrm{E}, g_\mathrm{I}) =  \frac{b_1}{a_1} = g_\mathrm{C} (E_\mathrm{E} - {\overline v}_\mathrm{som}) \,.
\end{aligned}
\end{align}\end{linenomath*}
In practice, the maximum attainable current for realistic conductance values is significantly smaller than $J_\mathrm{max}$, limiting the maximum firing rate. This must be taken into account when selecting the neuron tuning curve.

Model parameters can be estimated by solving the QP
\begin{linenomath*}\begin{align}
\begin{aligned}
&~ \min_{\substack{b_0,b_2,\\a_0, a_1, a_2}} \sum_{\substack{i,\\J_i \gg J_\mathrm{th}}} \left(b_0 + b_1 g_{\mathrm{E}}^i + b_2 g_{\mathrm{I}}^i - J^i a_0 - J^i g_{\mathrm{E}}^i a_1 - J^i g_{\mathrm{I}}^i a_2 - g^i_\mathrm{E} \right)^2 \,,
\end{aligned}
\label{eqn:h_optimal_parameters}
\end{align}\end{linenomath*}
subject to the nonnegativity constraints in \cref{eqn:h_model}, where $J^i = G^{-1} [ \mathscr{G}( g_{\mathrm{E}}^i, g_{\mathrm{I}}^i ) ]$. The conductances $g_\mathrm{E}^i$, $g_\mathrm{I}^i$ should be sampled over the operating range of the neuron, and samples with zero or very small output rates ignored: the inverse of $G$ is not defined for a zero output rate, and $H$ was derived under the assumption of superthreshold dynamics.

\subsubsection*{Synaptic weights as a quadratic program}
\label{sec:qp}
Given the nonlinearity model $H$ as defined in \cref{eqn:h_model}, our goal is to find weights $\vec w^E_i$, $\vec w^I_i$ such that a desired current $( \vec \jmath )_k$ flows into the soma for every sample $k$, cf.~\cref{eqn:decode_nonneg}.
Due to $H$ being a rational function, we cannot directly minimize the loss function in \cref{eqn:decode_nonlinear_synapses} as a quadratic program.
Instead, we solve a related quadratic program which, in practice, produces reasonable solutions. In particular, we simply equate the desired currents and the input-dependent nonlinearity model $H$. Let $\div$ and $\circ$ denote elementwise division and multiplication. Then, in vector notation, we get
\begin{linenomath*}\begin{align*}
\vec \jmath_i &= \frac{b_0 + b_1 A \vec w^E_i - b_2 A \vec w^I_i}
{  a_0 + a_1 A \vec w^E_i + a_2 A \vec w^I_i} \,, &\text{where } a_0, a_1, a_2, A, \vec w^E_i, \vec w^I_i \geq 0 \,,
\end{align*}\end{linenomath*}
and $A$ is a matrix of pre-population activities. Rearranging into a canonical form yields
\begin{linenomath*}\begin{align}
{{\vec \jmath}_i} \circ \left( a_0 + a_1 A \vec w^E_i + a_2 A \vec w^I_i \right) &= b_0 + b_1 A \vec w^E_i - b_2 A \vec w^I_i \,, \nonumber \\
\Leftrightarrow (a_1 {J_i} \circ A - b_1 A) \vec w_i^E + (a_2 {J_i} \circ A + b_2 A) \vec w_i^I &= b_0 - a_0 \vec \jmath_i \,, \text{ where } ({J_i})_{mn} = (\vec \jmath_i)_{m} \,, \nonumber \\
\Leftrightarrow \mathfrak{A}_i \vec w'_i &= \vec{\mathfrak b}_i \,,
\label{eqn:conductance_qp_matrices}
\end{align}\end{linenomath*}
where $\mathfrak{A}_i = (a_1 J_i \circ A - b_1 A, a_2 J_i \circ A + b_2 A)^T$, $\vec{\mathfrak b}_i = (b_0 - a_0 {{\vec \jmath}_i})$, and $\vec w'_i = (\vec w^E_i, \vec w^I_i)$.

We account for the subthreshold equality relaxation, by splitting $\mathfrak{A}_i$, $\mathfrak{\vec b}_i$ according to the samples invoking a zero firing rate (resulting in $\mathfrak{A}^-_i$, $\mathfrak{\vec b}^-_i$) and those invoking a positive firing rate (resulting in $\mathfrak{A}_i^+$, $\mathfrak{\vec b}_i^+$). Thus, the quadratic program becomes:
\begin{linenomath*}\begin{align}
\begin{aligned}
\min_{\vec w'_i, \vec s_i}  &~ \frac{1}2 (\vec w'_i)^T \left( (\mathfrak{A}^+)^T \mathfrak{A}^+ \right) \vec w'_i - ({\vec{\mathfrak b}}_i^+)^T \mathfrak{A}^+ \vec w'_i + \lambda N \|\vec w'_i \|_2^2 + \| \vec s_i \|_2^2 \,, \\
\text{subject to} &~ \, \mathfrak{A}_i^- \vec w'_i - \vec s_i \geq \mathfrak{\vec b}_i^- \text{ and } \vec w'_i \geq 0 \,.
\end{aligned}
\label{eqn:conductance_qp}
\end{align}\end{linenomath*}

\section{Experiments and Results}
\label{sec:experiments}

In this section, we validate the methodology described above. To this end, we perform three experiments. In Experiment~1, we test how well the two-compartment LIF nonlinearity model $H$ predicts simulation data for a single spiking neuron. In Experiment~2, we study the computational properties of $H$ under optimal circumstances when approximating random bandlimited functions. Finally, in Experiment~3, we test whether the results from Experiment~1 and 2 are still visible in the context of a noisy, feed-forward spiking neural network.

Unless explicitly specified, the neuron model parameters are chosen according to \Cref{tbl:parameters}.  We model fast excitatory synapses as an exponential low-pass with a time-constant of \SI{5}{\milli\second} as found in glutamatergic pyramidal neurons with AMPA receptor \citep{jonas1993quantal}. The inhibitory pathway is modeled with a \SI{10}{\milli\second} time-constant as found in inhibitory interneurons with GABA\textsubscript{A} receptors \citep{gupta2000organizing}. We use the \texttt{cvxopt} library \citep{vandenberghe2010cvxopt} as a QP solver for \cref{eqn:h_optimal_parameters}, and the \texttt{osqp} library \citep{osqp} to solve \cref{eqn:conductance_qp}.\footnote{\texttt{osqp} is used as a part of \texttt{libbioneuronqp}, see \url{https://github.com/astoeckel/libbioneuronqp}} The source code of the computer programs used to conduct and evaluate the experiments can be found in the supplemental materials.

\subsection{Experiment 1: Fitting the surrogate model to simulation data}
\label{sec:fit_model}

The surrogate model $H(g_\mathrm{E}, g_\mathrm{I})$ in \cref{eqn:h_model} predicts the current flowing into the somatic compartment of a two-com\-part\-ment LIF neuron for excitatory and inhibitory conductances $g_\mathrm{E}$ and $g_\mathrm{I}$. $H$ is characterized by five parameters $a_0$, $a_1$, $a_2$, $b_0$, $b_2$, for which a coarse theoretical estimate can be derived from \cref{eqn:h_model_prototype}.
We compare spike rates measured in numerical simulation to the spike rate prediction obtained before and after fitting the model parameters to empirical data using \cref{eqn:h_optimal_parameters}. We do this both for constant conductances $g_\mathrm{E}$, $g_\mathrm{I}$ and for artificial temporal spike noise superimposed on the conductance pair.

\subsubsection*{Experiment 1.1: Constant conductances}

We consider three two-compartment LIF neurons with different coupling conductances $g_\mathrm{C}$, namely $\SI{50}{\nano\siemens}$, $\SI{100}{\nano\siemens}$, and $\SI{200}{\nano\siemens}$, and measure their output spike rate for constant conductances $g_\mathrm{E}$, $g_\mathrm{I}$ on a grid. The conductance range has been selected such that the maximum firing rate is \SI{100}{\per\second}, and the spike onset approximately coincides with the diagonal of the resulting $g_\mathrm{E}$-$g_\mathrm{I}$-rate contour plot. We measure the steady-state firing rate by taking the inverse of the median inter-spike-interval over the simulation period. We compare these data to the current predicted by $H$ according to the theoretical and optimized estimated parameter sets. Parameter optimization is based on a training-set of \num{200} conductance pairs sampled with uniform probability from the conductance-range. The final prediction error is computed over all \num{10000} grid points.

\begin{figure}[p]
	\includegraphics[scale=0.9]{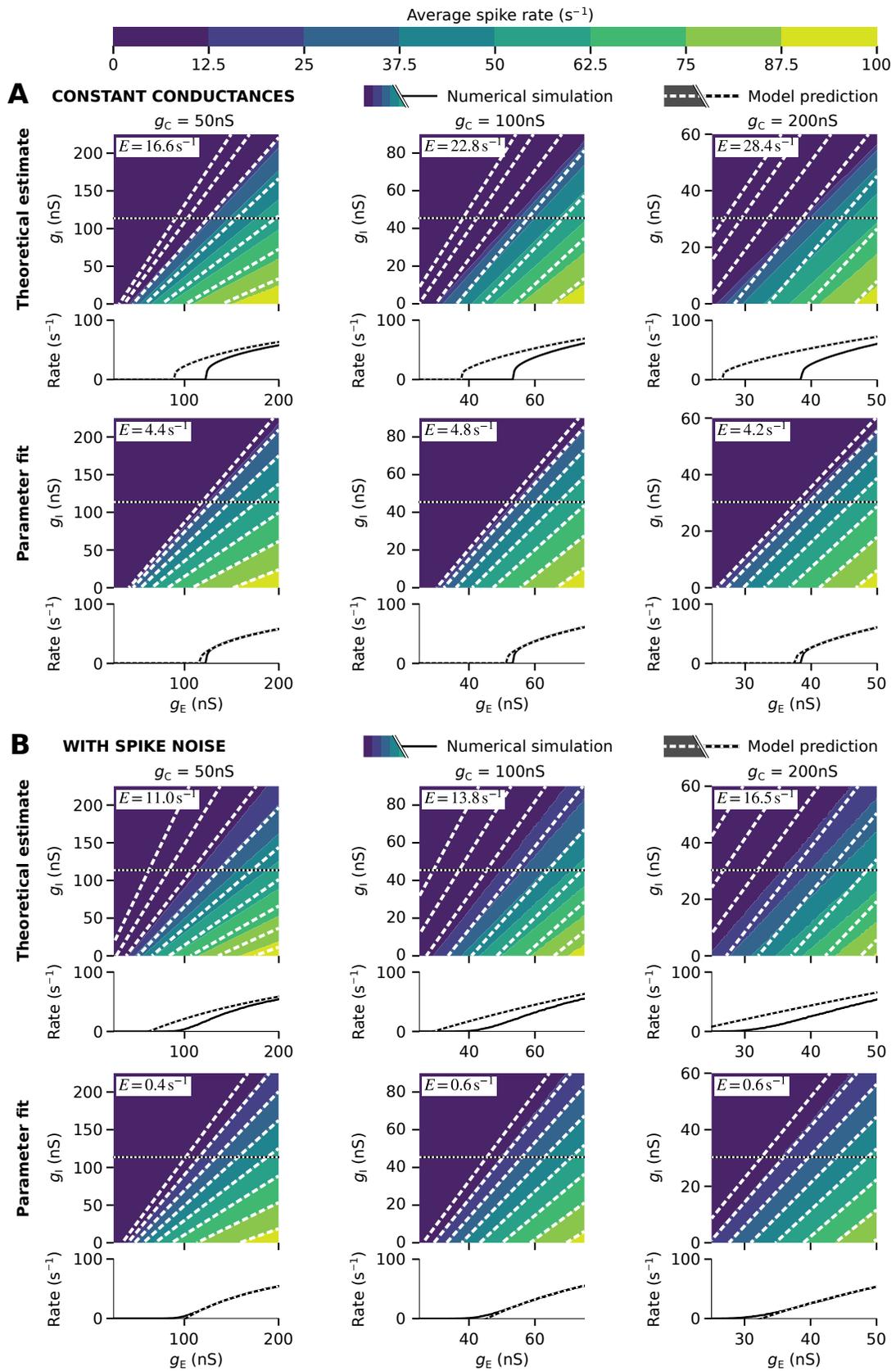}%
	\begin{subfigure}{0cm}\phantomcaption\label{fig:synaptic_nonlinearity_fit_a}\end{subfigure}%
	\begin{subfigure}{0cm}\phantomcaption\label{fig:synaptic_nonlinearity_fit_b}\end{subfigure}%
	\caption{Somatic current model parameter fits. (Continued on next page.)}
	\label{fig:synaptic_nonlinearity_fit}%
\end{figure}
\addtocounter{figure}{-1}
\begin{figure}
	\caption{Somatic current model parameter fits (see previous page). Average spike rates $\mathscr{G}(g_\mathrm{E}, g_\mathrm{I})$ measured in simulation are depicted as colored contour plots and solid lines in the cross-section.
		Dashed lines correspond to the model prediction $G[H(g_\mathrm{E}, g_\mathrm{I})]$. Dotted lines indicate the location of the cross-section.
		$E$ denotes the spike-rate RMSE over the regions where either the measured or predicted spike rate is greater than \SI{12.5}{\per\second}.
		Columns correspond to different values of $g_\mathrm{C}$.
		Top half shows the spike measured and modeled spike rates for constant conductance values $g_\mathrm{E}$, $g_\mathrm{I}$.
		The bottom half corresponds to conductance values with superimposed spike noise.
		The model fits the simulated response functions well after optimization.\\[-0.5cm]
		\rule{\columnwidth}{1pt}}
\end{figure}

The results for this experiment are depicted in \cref{fig:synaptic_nonlinearity_fit_a}, the fitted parameters can be found in \Cref{tbl:model_parameters}. When using the theoretical parameter estimate from \cref{eqn:h_model_prototype}, there is a significant discrepancy between the model prediction and the numerical simulation, especially for large $g_\mathrm{C}$. This discrepancy is greatly reduced after fitting the model parameters. The model prediction almost perfectly fits the empirical data for output spike rates greater than \SI{25}{\per\second}. However, it fails to predict the spike onset correctly, placing it too early with respect to increasing $g_\mathrm{E}$. Furthermore, the predicted slope at the spike onset is less steep than what is actually measured. As discussed in \cref{sec:somatic_current_model}, we can see the linearity of $H$ increase as $g_\mathrm{C}$ is increased, i.e., the contour lines are more \enquote{parallel} for larger $g_\mathrm{C}$. Still, with an overall Root Mean Square Error (RMSE) of about \SI{4}{\per\second}, the model provides a reasonable approximation of the empirical data.

\subsubsection*{Experiment 1.2: Conductances with artificial temporal spike noise}

In a network context, $g_\mathrm{E}(t)$ and $g_\mathrm{I}(t)$ are usually not constant, but instead modeled as the weighted sum of low-pass filtered spike trains, resulting in a considerable amount of \enquote{spike noise} being superimposed onto the signal. In this experiment, we simulate artificial spike noise as two Poisson spike sources (one excitatory, one inhibitory) with rate $1/\lambda = \SI{1800}{\per\second}$ for inhibitory synapses and a rate of $1/\lambda = \SI{4500}{\per\second}$ for excitatory synapses. These rates were obtained by fitting Poisson distributions to measurements from Experiment 3. Spikes arriving at different synapses are simulated by uniformly sampling a random weight from the unit-interval for each spike event. The time-averaged conductance equals $g_\mathrm{E}$, $g_\mathrm{I}$, respectively. Apart from the simulation period being extended to one hundred seconds, the remaining experimental setup is unchanged from the last experiment.

As can be seen in our results for noisy conductances in \cref{fig:synaptic_nonlinearity_fit_b}, and particularly in the bottom cross-sections of the measured spike-rates in \Cref{fig:synaptic_nonlinearity_fit_b}, the steep spike-onsets predicted by the theoretical LIF response curve $G[J]$ (eq.~\ref{eqn:lif_rate}) are no longer present. Furthermore, the relationship between $g_\mathrm{E}$ and the rate appears to be roughly linear in each cross-section of the $g_\mathrm{E}$-$g_\mathrm{I}$-rate plot, which is not well captured by the standard LIF response curve. Hence, a \enquote{soft} version of the LIF response curve that takes noise into account would be a better choice when fitting the parameters \citep{hunsberger2014competing,capocelli1971diffusion,kreutz2015mean}. We instead take a pragmatic approach and define $G$ as a rectified linear unit (ReLU), that is $G[J] = \max\{ 0, \alpha J + \beta \}$. Due to a relatively pronounced noise floor we only consider samples with a spike rate greater than \SI{12.5}{\per\second} for both fitting the model parameters and reporting the RMSE.

Fitting the parameters using the ReLU results in excellent accuracy for spike rates greater than \SI{12.5}{\per\second} (RMSE less than \SI{1}{\per\second}). However, the fitted model does not capture the subtle sigmoid shape of the response curve near the spike onset that is particularly pronounced for larger $g_\mathrm{C}$.

To summarize, the results indicate that, after fitting the model parameters, our surrogate model $H$ can indeed be used to predict the neural response curve with a relatively high accuracy. When taking noise in the input into account, it can be beneficial to choose a different neural response curve $G[J]$.

\subsection{Experiment 2: Computational properties of the two-compartment LIF nonlinearity $H$}
\label{sec:computational_properties}

The above experiment is encouraging in that our input-dependent nonlinearity $H$ seems to predict somatic currents well. Hence, in this experiment, we assume that $H$ accurately describes the somatic current and analyze whether there, in theory, could be any advantage to using a neuron model with this kind of input-depdendent nonlinearity.

Conceptually, what we would like to do is to measure how well a system can approximate functions of increasing complexity. If systems using input-dependent nonlinearities are able to approximate complex functions with a lower error than linear systems, they are \enquote{computationally more powerful}. Since \enquote{complexity} is somewhat ill-defined, we choose the spatial frequency content of a function as a proxy.

In our experiment, we randomly generate bandlimited current functions $J_\sigma(x, y)$ over the compact domain $(x, y) \in [-1, 1]^2$ on a $63 \times 63$ grid. $\sigma$ is a spatial low-pass filter coefficient that is inversely proportional to the bandwidth of the function. The functions we generate are normalized such that the mean is zero and the standard deviation equals $\SI{1}{\nano\ampere}$. We measure how well we can approximate this desired post synaptic current for a single post-neuron with a given input-dependent nonlinearity. The pre-neuron populations are as depicted in \cref{fig:network_c}. Two independent pre-populations with \num{100} neurons each represent $x$ and $y$, respectively, and project onto a single post-neuron.
The population tuning curves are randomly generated in each trial with a maximum firing rate between \num{50} and \SI{100}{\per\second} per neuron.
All pre-neurons project both inhibitorily and excitatorily onto the post-neuron. We measure the static decoding error, i.e., the difference between the target current function and the output of the surrogate model assuming ideal pre-population tuning curves without any dynamics. We consider dynamical simulation of a spiking \emph{network} in Experiment 3.

All synaptic weights are computed by solving the QP in \cref{eqn:conductance_qp} for 256 randomly selected training samples. For the conductance-based two-compartment LIF nonlinear input-dependent current function  $H_\mathrm{cond} = H$ we use the model parameters derived in the last experiment (\Cref{tbl:model_parameters}).  We emulate linear input currents $H_\mathrm{cur} = J_\mathrm{E} - J_\mathrm{I}$ in terms of $H$ by setting the model parameters to $b_0 = a_1 = a_2 = 0$, $a_0 = b_1 = 1$, $b_2 = -1$. The regularisation parameters were selected independently for each setup (\cref{fig:2d_regularisation_sweep}). The final error is the normalised RMSE (relative to the standard deviation of $\SI{1}{\nano\ampere}$ of $f_\sigma$) over all \num{3969} grid points. We add normal distributed noise (zero mean, unit standard deviation) to the pre-activities when computing the error to test how well the computed weights generalize.

As a further point of comparison, we include a \enquote{two-layer} neural network setup (\cref{fig:network_b}). The 200 pre-neurons are tuned to both input dimensions $(x, y)$, and not $x$ and $y$ independently. This setup should perform the best---but keep in mind that when modeling neurobiological systems, we may be constrained to pre-populations that do not exhibit the kind of multivariate tuning to the input signals we assume here.

\begin{figure}[p]
	\centering
	\includegraphics[scale=0.9]{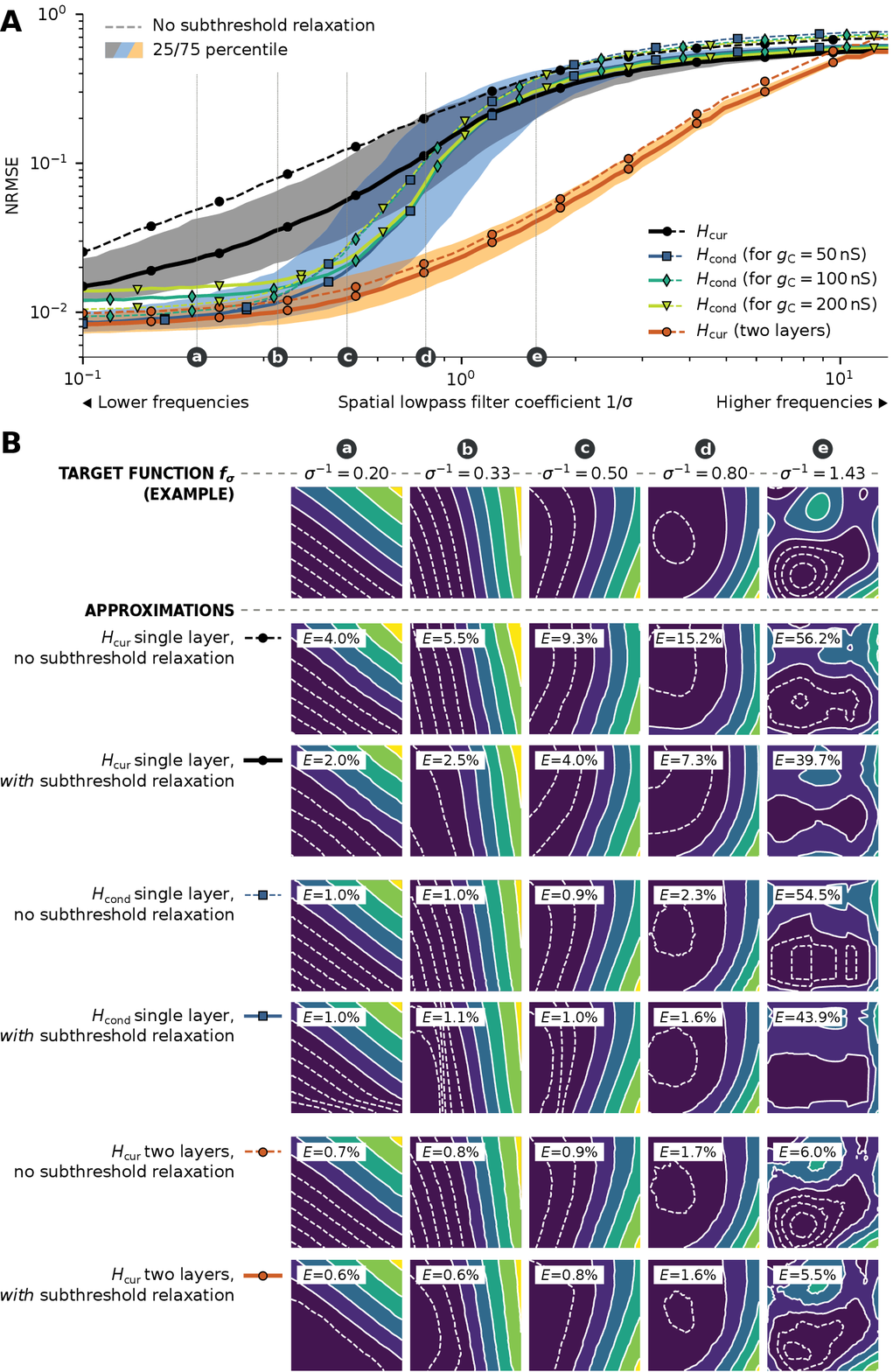}%
	\begin{subfigure}{0cm}\phantomcaption\label{fig:frequency_sweep_a}\end{subfigure}%
	\begin{subfigure}{0cm}\phantomcaption\label{fig:frequency_sweep_b}\end{subfigure}%
	\caption{Median decoding error for random functions. Continued on next page.}
	\label{fig:frequency_sweep}
\end{figure}
\addtocounter{figure}{-1}
\begin{figure}[p]
	\caption{Median decoding error for random multivariate current functions (see previous page). (A) Normalized RMSE between random, two-dimensional current functions and the decoded approximation using different input-dependent nonlinearities. The error measurement does not take subthreshold currents into account, using the function $\mathcal{E}$ defined in \cref{eqn:decode_current_subthreshold}. The low-pass filter coefficient $\sigma^{-1}$ is a proxy for the spatial frequency content in the target function. All points correspond to the median over \num{1000} trials. Dashed lines show results for not taking the subthreshold relaxation into account when solving for weights. The black lines show the results for a linear, current-based $H_\mathrm{cur}$; blue/green lines show the results for the two-compartment conductance-based model  $H_\mathrm{cond}$ with parameters given in \Cref{tbl:model_parameters} (without noise). Orange lines correspond a current-based network with two-dimensional pre-neuron tuning (i.e., a two-layer neural network). Shaded areas correspond to the 25/75 percentile for the current-based models and the conudctance-based model with $g_\mathrm{C} = \SI{50}{\nano\siemens}$. (B) Exemplary random functions and the corresponding decodings. Deep violet regions (dashed contour lines) correspond to subthreshold currents. Note how the shape of the subthreshold contour lines no longer match the target when subthreshold relaxation is active. As visible, the conductance-based nonlinearity $H_\mathrm{cond}$ helps to decode some target functions with a drastically smaller error compared to the current-based model $H_\mathrm{cur}$, especially when comparing the setups without subthreshold relaxation. $H_\mathrm{cond}$ does not provide any benefit for target functions with a high bandwidth.}
\end{figure}

Results are depicted in \cref{fig:frequency_sweep}. For a current-based neuron, and without subthreshold relaxation (dashed line in the plot), the median error increases linearly on a log-log plot from a $2.5\%$ error for low-frequency---almost linear---functions to an error of about $50\%$ for functions with a spatial cut-off frequency greater than one. Subthreshold relaxation reduces this error by up to $50\%$ for low $\sigma^{-1}$.

The error for the conductance-based nonlinearity increases sub-linearly on a log-log plot, starting at median errors of about $0.8\%$ for low-frequency functions. It is competitive with the two-layer network (see below) for $\sigma^{-1} < 0.5$. The error function converges to the results for the current-based model for spatial frequencies $\sigma^{-1} > 1$. Overall, the error for the conductance-based model is reduced by up to $65\%$ compared to the current-based model. The benefits of subthreshold relaxation are not as pronounced as for the linear current model.

The large errors for the current-based and the conductance-based models for $\sigma^{-1} > 1$ can be explained by the fact that both functions cannot be used to solve the XOR problem (cf.~\Cref{app:xor}). Functions with $\sigma^{-1} \geq 1$ are likely to possess multiple maxima/minima over $[-1, 1]^2$, akin to XOR, leading to a large error.

The two-layer network setup is able to approximate functions well up to $\sigma^{-1} = 10$, where it reaches the same final error values as the other setups. The complexity of the functions that can be approximated well by this setup is limited by the number of pre-neurons. The two-layer network can be thought of as linearly combining rectified hyperplanes to fit the target function, where each hyperplane is a single pre-neuron response. At a certain $\sigma$, the number of hyperplanes is no longer sufficient to reconstruct all the local maxima/minima in the target function.

To summarize, this experiment demonstrates that using a neuron model with the input-dependent nonlinearity $H$ significantly reduces the approximation error of current functions with $\sigma^{-1} < 1$ in a network setup with pre-populations independently representing the input dimensions. It is competitive with a two-layer network for $\sigma^{-1} < 0.5$.

\subsection{Experiment 3: Dendritic computation of multivariate functions in spiking network models}
\label{sec:dendritic_computation_network}

Experiment 1 suggests that we can use the non-linear post-synaptic current model $H$ to predict the average current flowing into the somatic compartment. Experiment 2 shows that we can, assuming that $H$ accurately describes the somatic current, approximate a somewhat larger class of random functions well. In our final experiment we study whether we can still observe the reduction in error in a feed-forward spiking neural network when using our model $H$ to solve for weights.

In contrast to previous experiment we do not base our error measurements on the decoded static somatic current for a \emph{single} post-neuron. Instead we decode the represented value from the neural activities of a target \emph{population} over time. Optimally, this decoded value should be $f(x(t), y(t))$, where $f$ is the function that we want to approximate and $x(t), y(t) \in [-1, 1]$ are input signals in turn represented by populations of \num{100} LIF neurons each. The target population either consists of standard current-based LIF neurons (\cref{fig:network_a}) or conductance-based two-compartment neurons (\cref{fig:network_c}).

Just as in the previous experiment, we consider the two-layer topology in \cref{fig:network_b} as a point of reference. Here, the input is mediated via an additional layer of \num{200} neurons representing the vectorial quantity $(x(t), y(t))$ over the interval $[-1, 1]^2$.

For all neurons, we generate tuning curves such that the maximum firing rate falls between \num{50} and \SI{100}{\per\second} over their represented range. Neurons are randomly marked as either excitatory or inhibitory. The probability of a neuron being inhibitory is 30\%. Excitatory and inhibitory synapses are modeled as a first-order exponential low-pass filter with time-constants of $\tau_\mathrm{E} = \SI{5}{\milli\second}$ and $\tau_\mathrm{I} = \SI{10}{\milli\second}$, respectively.

The network is simulated over \SI{10}{\second} at a time-resolution of \SI{100}{\micro\second}. Inputs $x(t)$ and $y(t)$ are sampled by moving through time along a fourth-order space-filling Hilbert curve over $[-1, 1]^2$. The output of the target population is decoded and filtered with a first-order exponential low-pass at $\tau = \SI{100}{\milli\second}$. We compute the desired target value $f(x, y)$ from the original input and pass it through the same series of low-pass filters as the spiking signals. We use the average synaptic time-constant of \SI{7.5}{\milli\second} to emulate the effect of the synaptic low-pass filters. Our final measure $E_\mathrm{net}$ is the normalized RMSE between the decoded output and the target values over time; the normalization is relative to the standard deviation of the target signal.

All synaptic weights are computed by solving the QP in \cref{eqn:conductance_qp}, with the same mapping of the current-based onto the conductance-based model as in Experiment 2. The regularization term $\lambda$ has been chosen independently for each neuron type and model parameter set such that the network error $E_\mathrm{net}$ is minimized when computing multiplication (cf.~\cref{fig:regularization_parameter_sweep}).

\subsubsection*{Experiment 3.1: Random bandlimited functions}

\begin{figure}[t]
	\centering
	\includegraphics[scale=0.9]{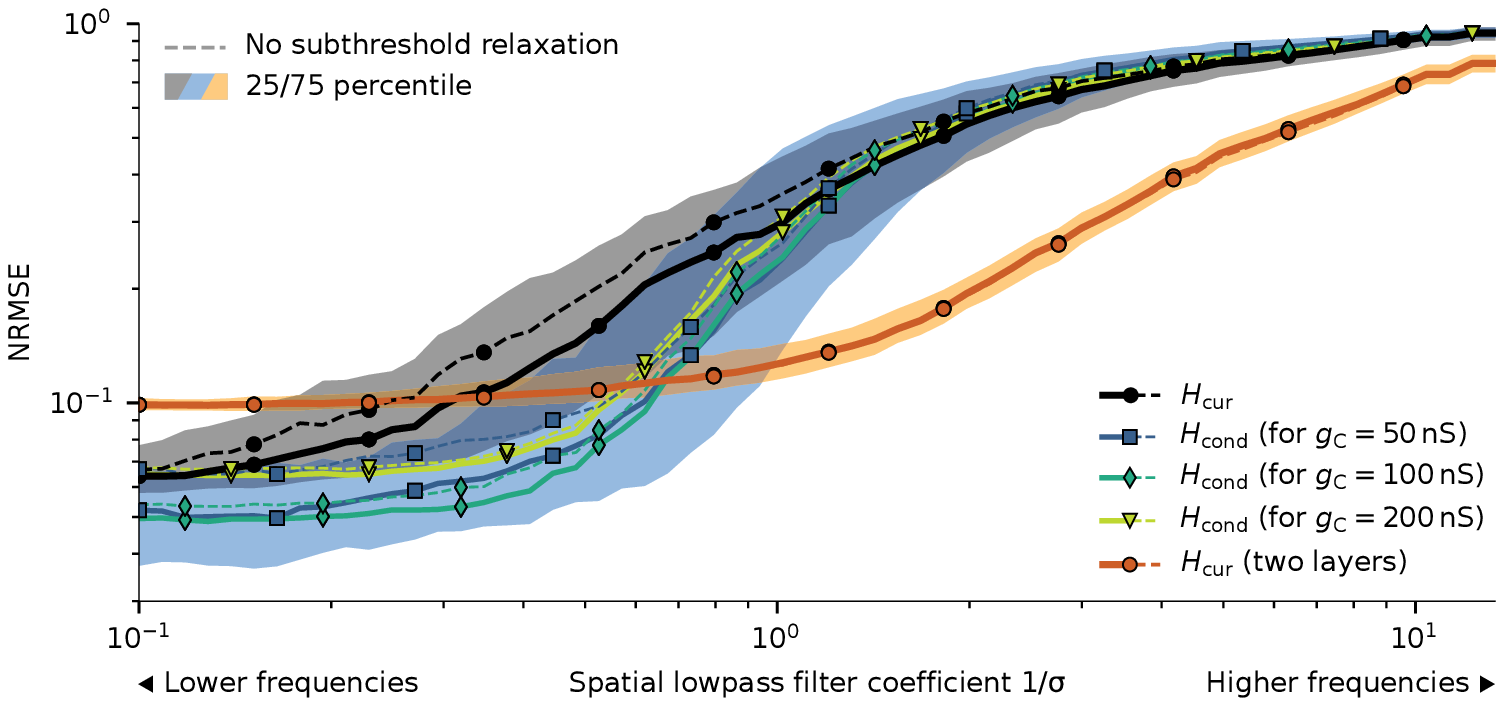}%
	\caption{Median error for computing random bandlimited functions in a feed-forward network over 1000 trials. Measured RMSE $E_\mathrm{net}$ is the difference between the represented value $D \vec a(t)$ and the expected value $f_\sigma(x(t), y(t))$ relative to the standard deviation of $f_\sigma$. See \cref{fig:frequency_sweep_a} for more detail.}
	\label{fig:frequency_sweep_network}
\end{figure}

We first test our network setup with the random bandlimited functions $f_\sigma$ we already used in the previous experiment. Results are depicted in \cref{fig:frequency_sweep_network}. Qualitatively, the results are very similar to what we saw before. The reduction in error between the current- and conductance-based models is not quite as large as suggested by the theoretical experiment, with a maximum reduction (in terms of the median) of only $45\%$ (instead of $65\%$ before). While subthreshold relaxation mostly increased the performance of the current-based model in the previous experiment, the improvement in error is now clearly visible for the conductance-based model as well.

Notably, the minimum median approximation error of the two-layer network is about $10\%$, whereas the single-layer current- and conductance-based models reach minimum errors of about $6.5\%$ and $5\%$, respectively. The two-layer network clearly surpasses the performance of the two-compartment LIF single-layer network for $\sigma^{-1} > 0.6$.
The larger errors are mainly caused by the representation of the two-dimensional quantity $(x(t), y(t))$ begin noisier than the representation of the scalars $x(t)$, $y(t)$ in the two pre-populations. This is because chaining multiple populations of spiking neurons slightly increases the noise floor. Furthermore, to cover the square $[-1, 1]^2$ as densely as two one-dimensional intervals $[-1, 1]$, we would optimally have to square the number of neurons. In our case, we would have to use \num{10000} instead of \num{200} neurons for the intermediate layer, which would not really be comparable to the single-layer setups---keep in mind that the two-layer network already uses $66\%$ more neurons.

\subsubsection*{Experiment 3.2: Benchmark functions}

\FloatBarrier

\begin{figure}
	\includegraphics[scale=0.9]{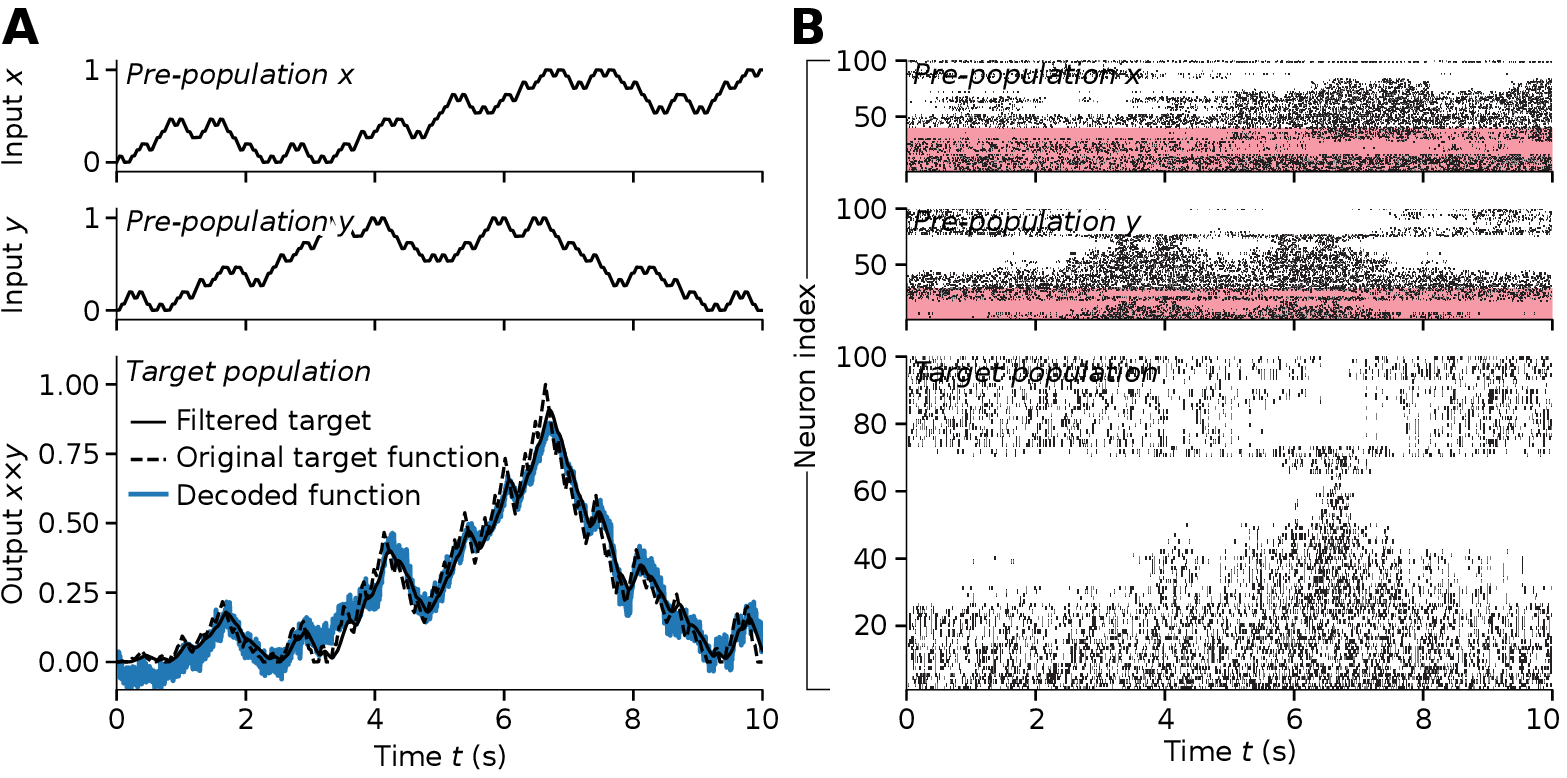}
	\caption{Single spiking neuron experiment showing computation of multiplication using a two-compartment LIF neuron. (A) Top two plots: inputs $x(t)$ and $y(t)$ as represented by the two pre-populations. The input is a fourth order 2D Hilbert curve. Bottom: mathematical target $f(x(t), y(t)) = x(t) y(t)$, filtered target function, as well as the decoded target population output. (B) Spike raster plots corresponding to the spiking activity of each of the populations. Red shaded background corresponds to inhibitory neurons in the pre-populations, all other neurons are excitatory.}
	\label{fig:spiking_example}
\end{figure}

\begin{table}[p]
	\caption{Spiking neural network approximation errors for function approximations on $[0, 1]^2$. Error values $E_\mathrm{net}$ correspond to the normalized RMSE (relative to the standard deviation of the target function) and are measured as the difference between the output decoded from the target population and the desired output for a ten second sweep across a 4th order 2D Hilbert curve over the input space. Results are the mean and standard deviation over 256 trials. The best result for a target function is set in bold; darker background colors indicate a worse ranking of the result in the corresponding row. Columns labeled \enquote{standard} refer to the default, single layer network setup, \enquote{two layers} refers to the two-layer setup, and \enquote{noise model} to the single layer network setup with model parameters derived under noise (see Experiment 1.2). Additional tables can be found in \Cref{app:tables}.}
	\centering\vspace{0.5cm}\scriptsize
	\begin{tabular}{r r r r r r r r }
		\toprule
		& \multicolumn{7}{c}{{\bf Experiment setup}} \\
		\cmidrule(l){2-8}
		& \multicolumn{3}{c}{
			LIF}
		& \multicolumn{2}{c}{
			Two comp. LIF $g_\mathrm{C} = \SI{50}{\nano\siemens}$}
		& \multicolumn{2}{c}{
			Two comp. LIF $g_\mathrm{C} = \SI{100}{\nano\siemens}$}
		\\
		\cmidrule(l){2-4}
		\cmidrule(l){5-6}
		\cmidrule(l){7-8}
		{\bf Target}
		& standard
		& standard\textsuperscript{\dag}
		& two layers\textsuperscript{\dag}
		& 
		standard\textsuperscript{\dag}
		& 
		noise model\textsuperscript{\dag}
		& 
		standard\textsuperscript{\dag}
		& 
		noise model\textsuperscript{\dag}
		\\
		\midrule
		$x + y$ 
		& \cellcolor{White!72!SteelBlue}4.2 $\pm$ 0.3\%
		& \cellcolor{White!58!SteelBlue}4.2 $\pm$ 0.3\%
		& \cellcolor{White!29!SteelBlue}8.2 $\pm$ 0.4\%
		& \cellcolor{White!100!SteelBlue}\bf2.3 $\pm$ 0.3\%
		& \cellcolor{White!43!SteelBlue}7.1 $\pm$ 0.8\%
		& \cellcolor{White!86!SteelBlue}3.7 $\pm$ 0.4\%
		& \cellcolor{White!15!SteelBlue}8.3 $\pm$ 0.9\%
		\\
		$x \times y$ 
		& \cellcolor{White!15!SteelBlue}26.6 $\pm$ 0.9\%
		& \cellcolor{White!29!SteelBlue}24.6 $\pm$ 0.9\%
		& \cellcolor{White!72!SteelBlue}9.2 $\pm$ 0.5\%
		& \cellcolor{White!86!SteelBlue}7.5 $\pm$ 1.1\%
		& \cellcolor{White!100!SteelBlue}\bf7.4 $\pm$ 1.3\%
		& \cellcolor{White!43!SteelBlue}10.9 $\pm$ 2.0\%
		& \cellcolor{White!58!SteelBlue}9.5 $\pm$ 2.0\%
		\\
		$\sqrt{x \times y}$ 
		& \cellcolor{White!15!SteelBlue}13.5 $\pm$ 0.6\%
		& \cellcolor{White!29!SteelBlue}12.5 $\pm$ 0.7\%
		& \cellcolor{White!43!SteelBlue}9.2 $\pm$ 0.4\%
		& \cellcolor{White!100!SteelBlue}\bf5.0 $\pm$ 0.8\%
		& \cellcolor{White!86!SteelBlue}6.2 $\pm$ 0.9\%
		& \cellcolor{White!58!SteelBlue}8.1 $\pm$ 1.7\%
		& \cellcolor{White!72!SteelBlue}7.9 $\pm$ 1.4\%
		\\
		$(x \times y) ^ 2$ 
		& \cellcolor{White!15!SteelBlue}45.6 $\pm$ 1.5\%
		& \cellcolor{White!29!SteelBlue}42.6 $\pm$ 1.5\%
		& \cellcolor{White!100!SteelBlue}\bf10.9 $\pm$ 1.1\%
		& \cellcolor{White!58!SteelBlue}19.7 $\pm$ 3.4\%
		& \cellcolor{White!86!SteelBlue}16.0 $\pm$ 3.4\%
		& \cellcolor{White!43!SteelBlue}22.4 $\pm$ 3.9\%
		& \cellcolor{White!72!SteelBlue}18.6 $\pm$ 4.1\%
		\\
		$x / (1 + y)$ 
		& \cellcolor{White!58!SteelBlue}5.6 $\pm$ 0.3\%
		& \cellcolor{White!72!SteelBlue}5.4 $\pm$ 0.3\%
		& \cellcolor{White!29!SteelBlue}8.1 $\pm$ 0.5\%
		& \cellcolor{White!100!SteelBlue}\bf2.3 $\pm$ 0.3\%
		& \cellcolor{White!43!SteelBlue}7.9 $\pm$ 1.2\%
		& \cellcolor{White!86!SteelBlue}3.8 $\pm$ 0.5\%
		& \cellcolor{White!15!SteelBlue}9.8 $\pm$ 1.5\%
		\\
		$\|(x, y)\|$ 
		& \cellcolor{White!43!SteelBlue}7.6 $\pm$ 0.5\%
		& \cellcolor{White!58!SteelBlue}7.4 $\pm$ 0.5\%
		& \cellcolor{White!29!SteelBlue}8.1 $\pm$ 0.4\%
		& \cellcolor{White!100!SteelBlue}\bf2.2 $\pm$ 0.2\%
		& \cellcolor{White!72!SteelBlue}6.4 $\pm$ 0.7\%
		& \cellcolor{White!86!SteelBlue}2.7 $\pm$ 0.4\%
		& \cellcolor{White!15!SteelBlue}8.9 $\pm$ 0.9\%
		\\
		$\mathrm{atan}(x, y)$ 
		& \cellcolor{White!43!SteelBlue}9.4 $\pm$ 0.5\%
		& \cellcolor{White!58!SteelBlue}9.0 $\pm$ 0.5\%
		& \cellcolor{White!29!SteelBlue}9.7 $\pm$ 0.5\%
		& \cellcolor{White!100!SteelBlue}\bf4.0 $\pm$ 0.8\%
		& \cellcolor{White!72!SteelBlue}7.4 $\pm$ 0.8\%
		& \cellcolor{White!86!SteelBlue}6.1 $\pm$ 1.2\%
		& \cellcolor{White!15!SteelBlue}11.6 $\pm$ 1.2\%
		\\
		$\max(x, y)$ 
		& \cellcolor{White!15!SteelBlue}14.9 $\pm$ 0.6\%
		& \cellcolor{White!29!SteelBlue}13.8 $\pm$ 0.6\%
		& \cellcolor{White!72!SteelBlue}8.4 $\pm$ 0.3\%
		& \cellcolor{White!86!SteelBlue}6.9 $\pm$ 0.7\%
		& \cellcolor{White!100!SteelBlue}\bf6.4 $\pm$ 0.7\%
		& \cellcolor{White!43!SteelBlue}9.4 $\pm$ 1.1\%
		& \cellcolor{White!58!SteelBlue}9.0 $\pm$ 0.7\%
		\\
		\bottomrule
	\end{tabular}\\[0.2cm]
	\raggedright\textsuperscript{\dag}With subthreshold relaxation
	\label{tbl:function_approximations}
\end{table}

While the random functions in the above experiments (see \cref{fig:frequency_sweep_b} for an example) are useful to systematically characterize the individual setups, it is hard to tell from these data alone what the practical impact of the two-compartment LIF neuron is. To this end, we selected eight mathematical benchmark functions $f(x, y)$ and repeated the experiment. Functions include the maximum $\max(x, y)$, and various forms of multiplication ($\sqrt{x \times y}$, $x \times y$, $(x \times y)^2$; see \cref{tbl:functions} for a complete list). Note that we compute all these functions over the interval $[0, 1]^2$ instead of $[-1, 1]^2$ by shifting and scaling the values represented by the neuron populations, i.e., we compute $f\big((x+1)/2, (y + 1)/2\big)$. As mentioned above and proved in \Cref{app:xor}, we know that we are not be able to solve the XOR problem with the two-conductance LIF neuron, and multiplication over $[-1, 1]^2$ can be thought of as a continuous form of XOR. We should be able to approximate multiplication over $[0, 1]^2$ one quadrant however.\footnote{The obvious solution to approximating \enquote{full} multiplication using two-compartment LIF neurons is to split the target population into four quadrants; however, we wanted to use network setups that are not optimized for a particular problem.}

A summary of the results over $256$ trials per function and setup is given in \Cref{tbl:function_approximations}, traces from an example trial are depicted in \Cref{fig:spiking_example}. More detailed results can be found in \Cref{tbl:function_approximations_complete}. For all but one target function (squared multiplication, which has the highest bandwidth of all tested functions), the conductance-based two-compartment model with a coupling conductance of $g_\mathrm{C} = \SI{50}{\nano\second}$ achieves the smallest error $E_\mathrm{net}$. Using the surrogate model parameters derived under noise is beneficial when computing multiplicative functions and the maximum. For these target functions, the synaptic connection matrix tends to be sparser, increasing the input noise. Apparently, this increase in noise matches the environment the neuron parameters have been optimized for. Interestingly, a purely current-based, single-layer network is competitive for all functions except for multiplication. The minimum error for the two-layer network is about $8\%$ even for simple functions, matching the observation we made in the random function experiment above.

An effect that could contribute to the superior performance of the two-compartment neuron model in some experiment are the low-pass filter dynamics of the dendritic compartment. These filter the high-frequency spike noise and thus may reduce the target error. We control for this effect in an experiment described in \Cref{app:pre_filter}, where we add an optimal low-pass filter to each network setup. Results are shown in \Cref{tbl:function_approximations_pre_filter}. We find that a matched pre-filter consistently reduces the error of all setups by only $1\%-2\%$, which indicates that the low-pass filter dynamics of the dendritic compartment are not the primary source for the reduction in error.

To summarize our experiments, we demonstrate in three stages (validation of the nonlinearity model $H_\mathrm{cond}$ for a single neuron, purley mathematical properties of $H_\mathrm{cond}$, and, finally, performance on a network-level) that we are able to successfully incorporate an---admittedly simple---model of nonlinear passive dendritic interaction into functional modeling frameworks. Instead of reducing the accuracy of our networks, the added detail can be systematically leveraged for computation. Our experiments also suggest that---at least in a biologically informed setting, i.e., using spiking neurons---this type of computation may result in a higher accuracy compared to two-layer architectures that suffer from an increase in the amount of spike-induced temporal noise due to the additional neuron layer.

\section{Discussion}
\label{sec:conclusion}

We derived a mathematical model of input-depdendent post-synaptic currents in a two-compartment LIF neuron that can be interpreted as a simple form of passive dendritic computation. We experimentally demonstrated that networks with fewer layers but biophysically plausible nonlinearities can compute a broad range of multivariate functions as well as or better than networks typically constructed using functional modeling frameworks. In particular, we proposed a mathematical model $H$ that captures nonlinear interactions between input channels, for example caused by conductance-based synapses or the dendritic tree. By mapping individual channel states onto an average somatic current $J$, this model can be integrated into mathematical frameworks that classically rely on current-based input channels.

Specifically, we demonstrated how to incorporate the dendritic nonlinearity $H$ into the Neural Engineering Framework (NEF). To this end, we discussed extensions to the NEF that allow us to optimize for nonnegative synaptic weights that invoke a desired somatic current $J$, and relax the optimization problem by taking subthreshold currents into account. We combined these methods with a specific surrogate model for $H$ in the context of a two-compartment LIF neuron. Finally, we performed a series of spiking neural network simulations that show that our methods allow dendritic nonlinearities to be systematically exploited to efficiently approximate nonlinear multivariate functions up to a certain spatial bandwidth.

While our approach is a step towards providing a general model of dendritic computation in top-down neurobiological modeling frameworks, it admittedly has several limitations. Most importantly, we treat the dendritic nonlinearity $H$ as time-independent. Correspondingly, we implicitly assume that synaptic time-constants typically dominate the overall neuronal dynamics. However, dendritic trees in biology---especially when considering active channels and dendritic spikes \citep{koch1999biophysics}---possess filter properties and adaptation processes that are not accounted for in our model. It would be interesting to incorporate the dynamical properties of dendritic trees into the NEF by employing the recent techniques presented by \cite{voelker2018improvinga}.

A further shortcoming of the derivation of the surrogate model of $H$ for the two-compartment neuron model is the assumption that the average somatic membrane potential is constant. While we are able to alleviate this assumption to some degree by fitting the model parameters to simulation data, the exact model parameters depend on the specific working-regime in which the neuron is used. Deviations from the modeled behavior are particularly apparent in situations with output firing rates smaller than ten spikes per second (cf.~\cref{fig:avg_vsom_no_ref,fig:synaptic_nonlinearity_fit}). Correspondingly, the dendritic nonlinearity presented in this paper may not be a suitable model for brain areas featuring extremely low maximum firing rates. There are two potential ways to work around this limitation. First, it may be possible to include an input-dependent membrane potential term in the nonlinearity. Or, second, one could directly use a sampled model for $H$. While these approaches are compatible with the concept of dendritic nonlinearity as introduced above, they both increase the mathematical complexity of the weight optimization problem to a point where strategies such as stochastic gradient descent are required. These techniques tend to have significantly weaker guarantees regarding finding an optimal solution compared to the convex quadratic programs employed in this paper.

In light of the above limitations, we would like to re-emphasize that, as stated in the introduction, our goal is not to provide a detailed mechanistic model of dendritic computation. Instead, we hope to provide a useful tool that captures essential aspects of dendritic computation---a nonlinear interaction between input channels---while being computationally cheap and mathematically tractable, but still grounded in biophysics. This helps to bridge the gap between purely abstract functional networks and more biophysically grounded mechanisms.

A potential application of our work outside of neurobiological modeling is programming neuromorphic hardware. Neuromorphic computers are inspired by neurobiological principles and promise to reduce the energy consumption of certain computational problems by several orders of magnitude compared to conventional computers \citep{boahen2017neuromorph}. Especially when considering mixed analogue-digital neuromorphic hardware systems, it should be possible to achieve a higher energy efficiency by implementing a more complex model neuron---such as the two-compartment LIF neuron discussed here---and performing local analog computation. Potential future work in this regard would be to validate our methods on a neuromorphic computing platform that implements dendritic trees, such as the \emph{BrainScales 2} system \citep{schemmel2017accelerated}.

Another line of future work is to consider arbitrary configurations of passive dendritic trees beyond the two-compartment LIF model. By applying Kirchhoff's circuit laws, any passive dendritic tree configuration can be described as a linear dynamical system. Correspondingly, it is possible to derive the dendritic nonlinearity $H$. It would be interesting to see whether it is still possible to relatively quickly optimize connection weights and in how far the number of compartments influences the computational power of the dendritic nonlinearity.

In conclusion, we believe that the methods proposed here provide a solid grounding for future work exploring both detailed biophysical mechanisms in the context of functional spiking networks, and improving neuromorphic methods for neural computation. We have shown how to cast the determination of connection weights in a functional network with conductance based synapses as an optimization problem with guaranteed convergence to the minimum. This optimization not only exploits known dendritic nonlinearities, but respects specifiable network topologies that conform to Dale's Principle. The result are functional spiking networks with improved accuracy and biophysical plausibility using fewer neurons than competing approaches.

\section*{Acknowledgments and Funding}

The authors would like to thank Aaron R.~Voelker for his comments on earlier drafts of this paper, as well as his advice and constructive criticism regarding the conducted experiments. This work was supported by the Canada Research Chairs program (no grant number), NSERC Discovery grant 261453, and AFOSR grant FA9550-17-1-002.


\end{document}